\documentclass[aoas]{imsart}

\RequirePackage{amsthm,amsmath,amsfonts,amssymb}
\RequirePackage[authoryear]{natbib}
\RequirePackage[colorlinks,citecolor=blue,urlcolor=blue]{hyperref}
\RequirePackage{graphicx}

\newtheorem{corollary}{Corollary}
\usepackage{mathrsfs}
\usepackage{mathrsfs,amsmath,amsthm,amssymb,color}
\usepackage{natbib,epsfig,graphicx,pdfpages}
\usepackage{rotating}
\usepackage{float}
\usepackage{bm}
\usepackage{caption}
\usepackage{subcaption}
\usepackage{ragged2e}
\usepackage[normalem]{ulem}
\usepackage{CJKutf8}

\usepackage{ragged2e}
\usepackage{algorithm}
\usepackage{algorithmic}

\def\n{\nonumber}

\def\beq{\begin{equation}}
\def\eeq{\end{equation}}
\def\beqr{\begin{eqnarray}}
\def\eeqr{\end{eqnarray}}
\def\beqrs{\begin{eqnarray*}}
\def\eeqrs{\end{eqnarray*}}
\def\bet{\begin{theorem}}
\def\eet{\end{theorem}}
\def\bel{\begin{lemma}}
\def\eel{\end{lemma}}
\def\bep{\begin{proposition}}
\def\eep{\end{proposition}}
\def\bg{\begin{figure}[tbph]\begin{center}}
\def\eg{\end{center}\end{figure}}

\def\bc{\begin{center}}
\def\ec{\end{center}}

\newtheorem{remark}{Remark}

\def\wt{\widetilde}

\def\bI{\mathbf{I}}

\def\mN{\mathcal{N}}

\def\mD{\mathcal D}

\def\mQ{\mathbb Q}
\def\mR{\mathbb{R}}
\def\mS{\mathcal S}
\def\mH{\mathcal H}
\def\mM{\mathcal M}

\def\mE{\mathcal E}

\def\mU{\mathcal{U}}
\def\mV{\mathcal{V}}

\def\mX{\mathbb{X}}

\def\mY{\mathbb{Y}}
\def\mZ{\mathbb{Z}}
\def\var{\mbox{var}}
\def\mJ{\mathcal J}
\def\mI{\mathcal I}

\def\mD{\mathcal D}
\def\mR{\mathbb{R}}
\def\mS{\mathcal S}

\def\mM{\mathcal M}
\def\mE{\mathcal E}

\newcommand{\ve}{{\varepsilon}}

\def\argmin{\mbox{argmin}}

\def\diag{\mbox{diag}}

\def\RSS{\mbox{RSS}}

\usepackage{array}
\newcolumntype{H}{>{\setbox0=\hbox\bgroup}c<{\egroup}@{}}

\startlocaldefs
\theoremstyle{plain}

\newtheorem{theorem}{Theorem}[section]
\newtheorem{lemma}[theorem]{Lemma}
\theoremstyle{remark}

\endlocaldefs

\begin{document}

\begin{frontmatter}
\title{SUPERVISED CENTRALITY VIA SPARSE NETWORK INFLUENCE REGRESSION: AN APPLICATION TO THE 2021 HENAN FLOODS' SOCIAL NETWORK}

\begin{aug}
\author[A]{\fnms{Yingying}~\snm{Ma}\ead[label=e1]{mayingying@buaa.edu.cn}},
\author[B]{\fnms{Wei}~\snm{Lan}\ead[label=e2]{lanwei@swufe.edu.cn}},
\author[C]{\fnms{Chenlei}~\snm{Leng}\ead[label=e3]{C.Leng@warwick.ac.uk}},
\author[D]{\fnms{Ting}~\snm{Li}\ead[label=e4]{tingeric.li@polyu.edu.hk}},
\and
\author[E]{\fnms{Hansheng}~\snm{Wang}\ead[label=e5]{hansheng@pku.edu.cn}},

\address[A]{School of Economics and Management,
Beihang University\printead[presep={,\ }]{e1}}

\address[B]{Center of Statistical Research,
Southwestern University of Finance and Economics\printead[presep={,\ }]{e2}}

\address[C]{Department of Statistics,
University of Warwick,\printead[presep={,\ }]{e3}}

\address[D]{Department of Mathematics,
The Hong Kong Polytechnic University,\printead[presep={,\ }]{e4}}

\address[E]{Guanghua School of Management,
Peking University\printead[presep={,\ }]{e5}}
\end{aug}

\begin{abstract}
The social characteristics of players in a social network are closely associated with their network positions and relational importance. Identifying those influential players in a network is of great importance as it helps to understand how ties are formed, how information is propagated, and, in turn, can guide the dissemination of new information. Motivated by a Sina Weibo social network analysis of the 2021 Henan Floods, where response variables for each Sina Weibo user are available, we propose a new notion of supervised centrality that emphasizes the task-specific nature of a player's centrality. To estimate the supervised centrality and identify important players, we develop a novel sparse network influence regression by introducing individual heterogeneity for each user. To overcome the computational difficulties in fitting the model for large social networks, we further develop a forward-addition algorithm and show that it can consistently identify a superset of the influential Sina Weibo users. We apply our method to analyze three responses in the Henan Floods data: the number of comments, reposts, and likes, and obtain meaningful results. A further simulation study corroborates the developed method.
\end{abstract}

\begin{keyword}
\kwd{Screening}
\kwd{Social networks}
\kwd{Spatial autoregression}
\kwd{Supervised centrality}
\end{keyword}

\end{frontmatter}

\section{Introduction}

Digitized social networks are ubiquitous. On social media such as Twitter, FaceBook and TikTok, communication becomes interactive and information is instantly shared, when users post texts, upload pictures and videos, and express opinions. Due to the growth of the internet and the growing availability of smart devices, these social networks are impacting all aspects of our lives including politics, economics,  and health.

The rapidly decreasing cost of collecting digitized information brings social networks that are increasingly
huge with millions or even billions of users/nodes. Despite their sizes, it is intuitively clear that not all the users
contribute equally to a network.  There is usually a subset of these users that exert major influence on determining
many aspects of the network including its formation, dynamics, and
topology \citep{motter2002cascade,kempe2005influential,centola2010the,zhao2017link,zhu2019portal,ma2021sparse}.
Because of this, identification of this subset of users has been a central research topic in network data analysis \citep{newman2010networks},
for which various centrality measures have been adopted.
{\color{black}Existing centrality measures can be broadly categorized into the following classes:
topology-based centrality \citep{freeman2002centrality, yang2018identifying}, topic-sensitive
centrality \citep{song2007identifying,cha2010measuring}, control centrality \citep{lin1974structural,cercel2014opinion},  and graph sampling recovery-based centrality  \citep{wei2019optimal,jin2023find}. These centrality measures have proven useful in a wide variety of applications. }
Because they rely exclusively on the topology of a network, they are less applicable when the influence is specific to certain responses or tasks.
For example,
a tweet announcement by the Twitter account of the American Statistical Association on a major
statistical meeting may be publicized differently than another on a piece of news on one of its
members, even though its network position and thus all network-dependent centrality measures
remain the same. That is, the influences of the American Statistical Association on disseminating
information regarding professional meetings and personal news are different.

This paper is empirically motivated by the problem of identifying influential social media users in a social network collected on Sina Weibo during the 2021 Henan Floods. In July 2021, Henan Province of China suffered severe flooding, sadly leading to 302 deaths and 50 people still missing. In the process of spreading local disaster rescue information, Sina Weibo,  one of the most popular social media in China, has played a crucial  role as an information dissemination platform. In a very short time after the disaster occurred, a number of posts about rescue information have been widely spread on this platform.
{\color{black} Knowing  the influence of each user that posts such rescue information can be beneficial for the propagation of rescue information \citep{ gruhl2004information, golbeck2006inferring}.

We review further the literature on centrality measures used to identify the most important users in a social network  \citep{FREEMAN1978215,wasserman1994social,sun2011survey}.  Some of the most traditional measures are degree, betweenness, harmonic, and eigenvector centrality. Degree centrality considers the number of connections each user has \citep{FREEMAN1978215}.   Betweenness centrality measures how often a user lies on the shortest paths between other users in the network  \citep{sun2011survey}.
Harmonic centrality measures the "average" distance of a user to other users, helping identify users that can quickly communicate with others \citep{marchiori2000harmony}.
Eigenvector centrality calculates a user's centrality as the weighted sum of its neighbors' centralities, with weights determined by the strength of connections between the user and its neighbors  \citep{HANSEN202031,jin2023find}.
}
{\color{black} Intuitively, these centrality measures rely exclusively on a binary view of the connections between users, where these connections represent an aggregate notion of relatedness, whether social, economic, or otherwise. It is not clear whether these centrality measures represent the best notion of importance when it comes to disseminating flood-related information. Given our interest in this specific type of information, the following fundamental question arises:
\begin{center}
Is there a way to identify and quantify influence triggered by a specific event?
\end{center}

To address this question, we aim to supplement the network topology with user-wise responses for the identification of task-specific influence. Specifically, we collected three variables: the number of reposts a Sina Weibo user obtained, the number of likes a Sina Weibo user received, and the number of comments other Sina Weibo users made on each user.  Intuitively, these variables are closely related to how information is spread, in addition to  network statistics. }
A well-connected user who does not repost is not likely to be influential in this network. Conversely, a poorly-connected user is unlikely to exert much influence on others, even if they frequently repost. For instance, within the context of the Henan Floods data, we observed that although the local media in Henan province extensively disseminated rescue information about the floods, it garnered limited attention due to their modest number of followers on Sina Weibo. In contrast, People's Daily and CCTV, two renowned media platforms with substantial followings, exerted significant influence only after many celebrities began sharing the flood rescue information these two platforms had posted on Sina Weibo. Specifically, over 25\% of the Sina Weibo users  in the data reposted flood-related content from these two media outlets.

The discussion above suggests that in many cases, influential users are often response-specific. To capture the dependence of the influence on a response variable, we define a response-specific influentialness as \textit{supervised centrality}. One of the uses of  supervised centrality is to accelerate important news propagation and
help  prevent the spread of    fake news \citep{gruhl2004information, vicario2019polarization}.  For example, for a target piece of fake news, we can collect  the   user  posting behaviors for such information     on  social network. The number of  reposts for each fake news spreader  can be used as the response variable.  Via  the supervised centrality method,   the influential user set for the spreaders can be identified. Sina Weibo users in the set are  critical to spread the fake news and
by eliminating such influential Sina Weibo users,  we  can  prevent about   80\% of the fake news spreading. This  is actually  evidenced by the  results of real data analysis in Table \ref{tab:network_effect}.

To estimate supervised centrality, we associate each user with its own influence parameter, and develop a novel sparse
network influence regression (SNIR) model  that
relates the userwise responses via the network structure with a sparsity assumption on the influence parameter.  Specifically in SNIR model,  the dependence between the Sina Weibo users
is captured by the response of a user depending on those of the adjacent ones,
in a way similar to the spatial autoregressive  (SAR) model in spatial analysis where the dependency is based on spatial proximity
\citep{ord1975estimation,cliff1975model}.  
The influential Sina Weibo users correspond to those with nonzero influence parameters. In the context of the Henan Floods dataset, that is, only those users with influence parameters different from zero impact the observed response of interest. Though this notion of influence can only serve as a crude approximation to the overall spectrum of influentialness, we have found it very useful in understanding the information flow in our chosen data analysis.
To estimate the influence parameters,
 one approach is to use the standard quasi-maximum likelihood method which turns out to be computationally very expensive for large networks that are the primary target of applications of this work.  
 As an alternative, we propose a novel and efficient estimation approach based on conditional quasi-maximum likelihood, after realizing that the number of influential Sina Weibo users is usually small in practice, making their contribution to the full likelihood negligible.
 To identify the set of those influential Sina Weibo users,
 we develop a  forward-addition algorithm and show that with high probability, the algorithm can select a superset of the influential Sina Weibo users.
The theoretical properties of the conditional quasi-maximum likelihood method and forward-addition algorithm
 are rigorously studied under appropriate  conditions.

We have applied our model to the Henan Floods network and identified three sets of influential users who helped to spread important rescue information based on reposts, comments and likes, respectively. These three influential user sets share some common users but also have influential users unique to each response, suggesting that  supervised centrality does depend on the response.
To validate the power of our approach for identifying influential users in this network, we compare its performance with those purely topology-based approaches and that of a purely response-based approach.
  The results are summarized in  Table \ref{tab:network_effect} in Section \ref{sec:data}.  {\color{black} We can see that the removal of the influential users identified by our approach has the largest impact on the
 network information spreading
 in terms of
 the amount of responses lost.}
 The comparison  also confirms that our approach can identify users that influence the network topology and the response simultaneously rather than separately.

The structure of this article is organized  as follows: In Section \ref{sec:data1}, we briefly describe the Henan Floods network dataset.
 We introduce the SNIR model, and discuss the estimation of its parameters and selection of the nonzero influence coefficients in Section \ref{sec:method}.
In Section \ref{sec:data}, we present the analysis of  Henan Floods data. A simulation study  is conducted  in Section \ref{sec:simulation}.
Concluding remarks are found in Section \ref{sec:discussion}. All technical details and proofs are relegated to the Appendix, which also includes  additional results on real data analysis. Specifically,
Appendix \ref{sec:dynamic model} explores an extension of the model  leveraging the dynamics of the data over time,
Appendix \ref{sec:choice} examines the impact of the choice of the candidate set $\mM$, and
Appendix \ref{sec:withX} extends the model by including exogenous variables  and  heterogeneous random errors.


\section{Henan Floods Social Network} \label{sec:data1}

On July 20, 2021, Henan province in China suffered severe flooding caused by heavy rain. By August 2, 2021, floods and secondary disasters have  affected millions of people   in Henan province, with 302 people killed and 50 missing. Official data showed that the average daily rainfall during this period in Zhengzhou, the provincial capital city of Henan, exceeded the historical maximum value since record began, with its average hourly rainfall exceeding even the daily historical extreme value. Roughly speaking, the amount of three day's rainfall was  equivalent to that of a year's average value.

During this major natural disaster, a wide range of information and news outlets have been mobilized to cover the event. In the process of disseminating local disaster  rescue information, social media, especially Sina Weibo, the Twitter counterpart in China, have played an important role.  In the morning of July 22 alone, the top 3 popular topics on Sina Weibo were all related to the flood, with each topic having received more than 100 million views, reaching 570 million at the highest. Several major news agencies in China such as ``People's Daily'' and ``CCTV''  have  posted  news on Sina Weibo and provided advices on how to support flood victims.
Among all the original rescue messages posted on Sina Weibo, 8    messages have  been repeatedly reposted by the majority of the celebrities. On Sina Weibo, one achieves the celebrity status by accumulating a large number of followers usually; see https://www.5566.net/ weibo.htm for ranked lists. Among the 8 messages, 3 of them are posted by  ``People's Daily'' while the other 5 messages are posted by ``CCTV news'', ``Sina Henan"",  ``Henan TV", ``Vientiane News'',  and ``Headline News'', respectively.  The largest number of total reposts is a message by ``People's Daily'' with
 more than  18,000,000 reposts. Among the 8 messages, 4 have received more than one million reposts while the other 4 each has hundreds of thousands  reposts (i.e., 168000, 230000, 401000, 471000 reposts, respectively).

\begin{figure}
     \centering
     \begin{subfigure}[b]{0.45\textwidth}
         \centering
         \includegraphics[width=\textwidth]{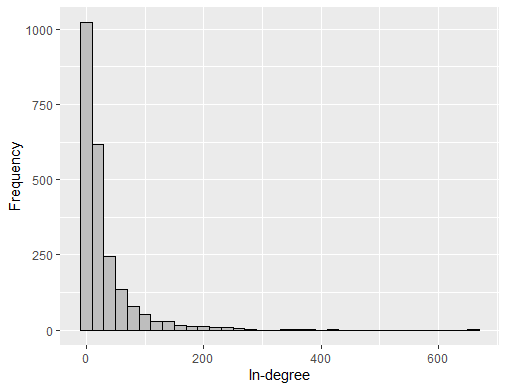}
         \label{fig:indegree_hist}
     \end{subfigure}
     \hfill
     \begin{subfigure}[b]{0.45\textwidth}
         \centering
         \includegraphics[width=\textwidth]{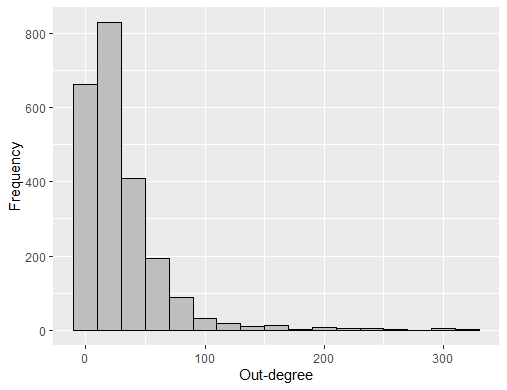}
         \label{fig:outdegree_hist}
     \end{subfigure}
     \caption{The histogram of in-degrees (left) and out-degrees (right) for  Henan Floods network.}
     \label{fig:degree_hist}
\end{figure}

 If a message on Sina Weibo has been forward by a large amount  of users. There will
 be a hot repost list under the original message.
  Usually, if the user has been verified by Sina Weibo as a
 celebrity, when  the user forward the message, this repost  will be
labeled as a hot repost.
Other reposts under this message can also be labeled as hot reposts  as long as they satisfy the following two conditions.
 First, the number of the   reposts/comments/likes for their
 retweet is sufficient large (e.g., larger than 100 for at least one criteria).
 Second, the popularity level of their retweet is ranked  among the top 500 repost users (note that Sina Weibo has a score to measure the popularity level based on
 number of repost, comments and likes on the retweet).
 When  the retweet satisfies such characters,  it   will be labeled as hot repost.
 Based on the hot repost information, we can identify a  hot repost user list.
 For example, about 477 hot repost users can be captured under one of the message posted by ``People's Daily''.  Based on the hot repost user list labeled under  the 8 original  messages, we can  identify 2275 Sina Weibo users, and over 90\% of them are celebrities  with Sina Weibo  certification.
 For these users, we further  extract their social relationship  on Sina Weibo, which can form a network structure $A=(a_{ij})$  {\color{black} and we assume $A$ is fixed. Specifically, if the $i$th user follows the $j$th user on Sina Weibo, we have $a_{ij}=1$, otherwise, $a_{ij}=0$. }
The resulting network has $\sum_{i,j} a_{i,j}=66,993$  edges and  $\sum_{i<j} a_{i,j}\bI(a_{i,j}a_{j,i}\neq 0)= 17,680$
mutual connected pairs,  where  $\mathbf{I}(\cdot)$ is an indicator function.
The network density (ND= $\sum_{i,j} a_{i,j}/N^2$) is around 1.3\%. We first conduct some preliminary data analysis by plotting the histograms of the in-degrees (the number of followers of each user) and out-degrees (the number of followees of each user) in Fig \ref{fig:degree_hist}. We can see the highly skewed  in-degree and out-degree distribution. Specially, the in-degree histogram is much more skewed than out-degree, which suggest that $A$ might be a power-law type network structure.
For such kind of network,    the size of  influential user set should be  limited,
which inspire us to propose a sparse assumption  on the influence parameter.

\begin{figure}
     \centering
     \begin{subfigure}[b]{0.3\textwidth}
         \centering
         \includegraphics[width=\textwidth]{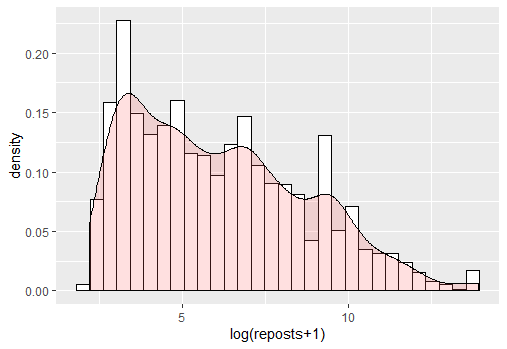}
         \caption{Density of reposts.}
         \label{fig:den_reposts}
     \end{subfigure}
     \hfill
     \begin{subfigure}[b]{0.3\textwidth}
         \centering
         \includegraphics[width=\textwidth]{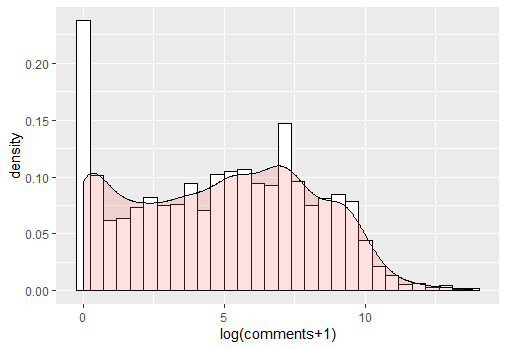}
         \caption{Density of comments.}
         \label{fig:den_comments}
     \end{subfigure}
     \hfill
     \begin{subfigure}[b]{0.3\textwidth}
         \centering
         \includegraphics[width=\textwidth]{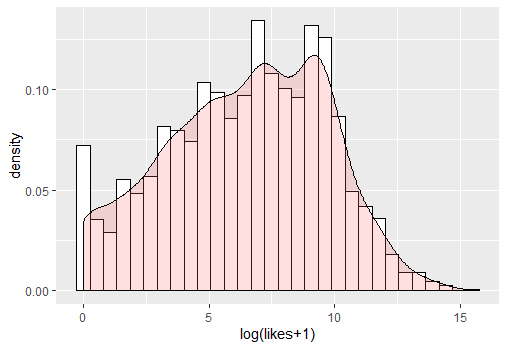}
         \caption{Density of likes.}
         \label{fig:den_likes}
     \end{subfigure}
        \caption{Density plot of  reposts, comments  and likes after logarithm transformation.}
        \label{fig:density0}
\end{figure}

Similar to Twitter, Sina Weibo allows one to repost, comment, or like (favorite) on a message, which provides  three natural measurements  for the response variable.
Specifically, the number of  reposts and likes  reflect how many times a message has been spread to other users, while the number of comments more or less reflects its popularity level \citep{siersdorfer2014analyzing}.  {\color{black} To incorporate their information in defining supervised centrality, we tally the logarithm of the number of reposts, comments, and likes of the 8 messages for the 2275 users as our response variables.}
We present in Fig \ref{fig:density0} their histograms and density plots. As we can see, the values of the three variables are relatively spread out.

\section{
Sparse Network Influence Regression
} \label{sec:method}

 The following notations are used throughout the paper. For a  vector
 $\bm u=(u_1,..., u_p)^\top,$ we denote  $\|\bm u\| = (\sum_{i=1}^{p} u_i^2)^{1/2} $
as its $l_2$ norm and  $\|\bm u\|_1 = \sum_{i=1}^{p} |u_i|$ as its $l_1$ norm.  For a matrix $A \in \mathbb{R}^{n\times n}$, we use $\lambda_{\max}(A) $
 and $\lambda_{\min}(A)$ to denote its largest and smallest eigenvalue, respectively. We further use $|A|$ to denote the determinant of the matrix.
 Finally, for a matrix
 $B=(b_{ij})\in \mathbb{R}^{m\times n}$, we denote $\|B\|_F=(\sum_{i,j}b_{ij}^2)^{1/2}$ as its Frobenius norm, and  $\|B\|_2=\sqrt{\lambda_{\max} (B^\top B) }$ as its operator norm. Denote  $\mS$ be a  candidate set and $\mS\backslash  j$ represents this
set  with the element  $j$ removed. For  a vector $\mV=(v_1,\cdots,v_N)^\top\in\mR^N$,
we define $\mV_{(\mS)}=(v_i:i\in\mS)\in\mR^{|\mS|}$ as the sub-vector of $\mV$ according to  $\mS$.

\subsection{The Model}\label{model}

 Motivated by the Henen Floods' data, we consider the modelling of directed networks in this paper.
For a directed network with $N$ Sina Weibo users indexed as $ 1\leq i\leq N$, we denote its adjacency matrix as $A = (a_{ij}) \in \mathbb{R}^{N\times N}$,
where  {\color{black}  $a_{ij} = 1$ if user $i$ in the Sina Weibo network follows user $j$ and $a_{ij} = 0$ otherwise.}
Following the convention,  we assume absence of self-loops, so that $a_{ii}=0$ for any $1\leq i\leq N$. Denote the response on user $i$ as $Y_i\in \mR$. Our observations consist of $A$ and $(Y_1, \cdots , Y_N)^\top$. To relate the response to the network structure in a meaningful way, we propose the following
Sparse Network Influence Regression
(SNIR) model
\beq
\label{ssar}
Y_i =\mu_i+\sum_{j = 1}^N \rho_j a_{ij} Y_j + \varepsilon_i, \quad i=1,\cdots, N,
\eeq
where $\mu_i$ is an intercept,
$\rho_j$  is the influence parameter for user $j$, and $\varepsilon_i$s are independent and identically distributed random errors. This model assumes that the response of a user depends on those users that it follows, but the extent of the influence that those users exert is vertex-specific.
For simplicity, we assume  that $\varepsilon_i$s are i.i.d. sub-Gaussian random variables with mean 0 and  an unknown variance  $\sigma^2$. Note that our result can be readily extended to the case that $\varepsilon_i$s having heterogeneous variance, as can be seen from Appendix \ref{sec:withX}.  To distinguish influential users from non-influential ones, we assume that the coefficient vector $\bm{\rho}=(\rho_1,\cdots,\rho_N)^\top\in\mR^{N }$ is sparse with the size of its support potentially much smaller than $N$. Obviously, the parameters in \eqref{ssar} are not identifiable. To overcome this, we further assume that $\mu_i=\mu$ if $\rho_i\not= 0$, and $\mu_i=0$ if $\rho_i= 0$. The latter means that the response of a non-influential user is fully dependent on the responses of the influential users. Thus, to identify those users that are influential implied by our model, we just need to identify those users with nonzero $\rho$ parameter.
 \begin{remark}
 As seen from \eqref{ssar}, $\bm{\rho}$ and $A$ offer complementary views on the importance of users in the presence of a response. Degree centrality, as measured by the total number of neighbors of a user, provides a binary qualitative view of its importance, but not its extent. Under the usual network data settings where only an adjacency matrix corresponding to the network is observed, this binary view is all that is available. However, in settings where, in addition to this network, one also observes user-wise responses, as in the Henan Floods data, we may leverage the information inherent in these responses to define a more refined centrality measure that depends on the particular response of interest. From this perspective, supervised centrality provides an additional quantitative assessment of this importance.

 The form of \eqref{ssar} also suggests the use of degree centrality for pre-selecting the support of $\bm{\rho}$. This is  natural, as we would expect information (influence) flows only when there is a connected link between two users. As argued above, supervised centrality contributes an additional quantitative measure of a user's influence, having realized that a link (which contributes as part of the degree centrality) must exist a priori for the corresponding user to exhibit influence at all. Finally, by allowing $\rho$ to be user-dependent, we aim to tease apart the influence heterogeneity,  the central problem this paper tackles.

 It appears that the form of  \eqref{ssar}  is similar to structural equation modeling.  In the current setup, however, we only have one observation for each user with possible endogeneity as in model  \eqref{ssar},  rendering the latter not applicable.
\end{remark}

\begin{figure}[htb]
	\begin{subfigure}{0.5\linewidth}\centering
\includegraphics[width=0.7\linewidth]{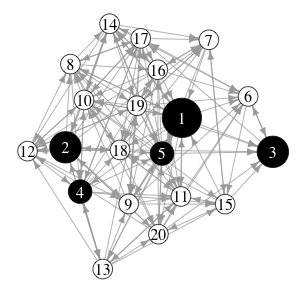}
\caption{Observed network}
\end{subfigure}
	\begin{subfigure}{0.5\linewidth}  \centering
\includegraphics[width=0.7\linewidth]{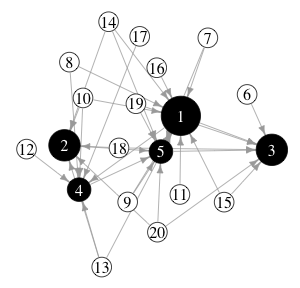}
\caption{Simplied network after removing links to non-influential users}
 \end{subfigure}
 \caption{A hypothetical network for SNIR model. The size of each user is proportional to its in-degree. Black users are influential and white users are non-influential. }
 \label{fig:ssar}
 \end{figure}

To appreciate the effect that SNIR method brings to network modelling, we examine a hypothetical network in Fig \ref{fig:ssar} with five influential users. The original observed network is plotted on the left, where a large number of connections are present. The plot on the right shows a simplified network with the links to non-influential users all removed.  We can see that by constraining the size of the support of $\bm{\rho}$, SNIR model can drastically reduce the complexity of the network structure for modelling purposes.


{\color{black}
 Spatial autoregressive type of models similar to \eqref{ssar} have been applied to network data in recent years because of the ease of changing the notion of neighbours from a spatial domain due to physical distance to a network domain due to connections \citep{leenders2002modeling,lin2010identifying,zhou2017estimating,QU2021180}. However, the SAR models employed in the existing literature are different from our SNIR model in several  important aspects. First,
 existing SAR models that can deal with a single realization
of the response variable often assume $\rho_i=\rho$,  that is, all the influence measures are the same. As a result, all the users that a user connects to have an equal influence. Such a parametrization is useful for capturing the average  dependence due to the network structure, but cannot distinguish influence for each individual user.
Second,   traditional SAR and  our SNIR model  have different influential mechanisms.  In the SAR model,  since the adjacency matrix is usually row-normalized, for a user to affect another individuals' behavior,  it needs to
affect most users in the network, otherwise the network effect coefficient
 (i.e., $\rho$) might not be significant \citep{liu2014endogenous}.  In contrast, the adjacency matrix in  SNIR model is not row-normalized and the influence of the influential  users in this model is aggregated to a certain sense. From an econometric point of view,
 SNIR method naturally incorporates the so-called social multiplier effect  and assumes that the social network is driven by the   influential users.
Because of the social multiplier effect,  one can still have significant influence   to the whole network with relatively few  influential users.
Note that in some econometric literature, the classical SAR and our SNIR model belong to  the local-average
and the local-aggregate model, respectively \citep{Liu2010,LIU2013243,lewbel2021social}, with the main difference  in whether the adjacency matrix is row-normalized or not.  Our SNIR model can be seen as a natural extension of the classical   local-aggregate model by allowing heterogeneous
 influence parameters
 in $\rho_j$s.
}

\subsection{ General estimation approach}\label{QMLE}
Before providing our strategy for estimating
the parameters when the support of $\bm \rho$ is unknown, we first introduce the notations used.
Denote  $\mY=(Y_1,\cdots,Y_N)^\top\in\mR^{N}$,   $\bm \mu=(\mu_1,\cdots, \mu_N)^\top\in\mR^{N}$,
$\mD=\mbox{diag}(\rho_j: $ $ 1\leq j\leq N)\in\mR^{N\times N}$ which  is a diagonal matrix, and
$\mE=(\varepsilon_1,\cdots,\varepsilon_N)^\top\in\mR^N$.
Define $\mS_F=\{1,2,\cdots,N\}$ as the index set of all the users,  $\mS_{1}=\{j: \rho_j\not=0, j \in \mS_F\}$ as the index set of the influential users, and $\mS_0=\mS_F\backslash\mS_{1}=\{ j: j\not\in\mS_{1},  j \in \mS_F\}$ as that of the non-influential ones. Let $\mN_i = \{j: a_{ji}\neq 0,  j \in \mS_F \}$ be the set of users that follows user $i$. For a subset $\mS$ of $\mS_F$, denote its cardinality as  $|\mS|$.  For a matrix  $M \in \mR^{N\times N}$ and two sets $\mS_a\subset\mS_F$ and $\mS_b\subset\mS_F$, we define $M_{(\mS_a,\mS_b)}$ as a sub-matrix of $M$ with its row indices in $\mS_a$ and column indices in $\mS_b$. Let $\mD_{(\mS_a)} = \mbox{diag}( \rho_j:j\in \mS_a )$ be a   sub-diagonal matrix of $\mD$ consisting of the elements with indices in $\mS_a$.

We can
rewrite \eqref{ssar} in matrix form as
$\mY =\bm \mu+ A \mathcal{D} \mY + \mE$
and immediately obtain the following log-likelihood function
\begin{equation*}
\label{loglik}
\ell(\mD,\bm \mu,\sigma^2; \mY, A) = -\frac{N}{2} \ln (2\pi \sigma^2) - \frac{1}{2 \sigma^2} (\mY-  H^{-1} \bm{\mu} )^\top H^\top H(\mY- H^{-1}\bm\mu) + \ln |H|,
\end{equation*}
where $H = \mI_N - A\mD$ and $\mI_N$ is an $N\times N$ identity matrix.
Since the support of $\bm\rho$ is not known, one approach to obtain its estimate is to maximize $\ell(\mD,\bm \mu,\sigma^2; \mY, A)$ with an $\ell_0$ constraint on  $\bm\rho$, subject to the additional constraint that $H$ is positive definite. This strategy gives rise to an NP-hard problem and thus is not computationally feasible. Another approach is to apply a convex relaxation of the $\ell_0$ penalty for example by employing a penalized likelihood method with an $\ell_1$ penalty on $\bm\rho$. It is not clear either how this approach works in practice due to the involvement of the parameter $\bm \rho$ in the term $ \ln |H|$ and the requirement on $H$ as a positive definite matrix, especially for large networks.

Because of this concern, we
  consider  a conditional maximum likelihood  estimation   approach. In order to illustrate this  approach more clearly,   we start with  the case where
 the  support of $\bm\rho$ is known before discussing  the unknown support case.
Write model \eqref{ssar} as:
\beqr
\mY_{(\mS_0)} &=& \bm 0_{(\mS_0)}+A_{(\mS_0,\mS_{1})} \mD_{(\mS_{1})}  \mY_{(\mS_{1})}  + \mE_{(\mS_0)} \label{ssar1}\\
\mY_{(\mS_1)} &=& \mu \bm 1_{(\mS_1)}+ A_{(\mS_1,\mS_{1})} \mD_{(\mS_{1})}  \mY_{(\mS_{1})}  + \mE_{(\mS_1)},\label{ssar0}
\eeqr
where $\bm 0_{(\mS_0)}$ is a vector of $0$'s and $\bm 1_{(\mS_1)} $ is vector of $1$'s.
Thus, $\bm \rho_{(\mS_{1})}$ can be estimated by minimizing the following least squares type objective function,
\beqr
\wt Q(\mu, \bm \rho_{(\mS_{1})}) &=&\|\mY_{(\mS_0)}- A_{(\mS_0,\mS_{1})} \mD_{(\mS_{1})}  \mY_{(\mS_{1})} \|^2 +(\mY_{(\mS_1)}- H_1^{-1}  \mu \bm 1_{(\mS_1)})^\top H_1^\top H_1 \n\\ &&(\mY_{(\mS_1)}-H_1^{-1}  \mu \bm 1_{(\mS_1)})  + 2\sigma^2\ln |H_1|   \n
\eeqr
with $H_1 = \mI_{|\mS_{1}|} - A_{(\mS_1,\mS_{1})} \mD_{(\mS_{1})}$.
The first term of $\wt Q(\mu, \bm \rho_{(\mS_{1})})$ is the conditional log-likelihood for $\mY_{(\mS_0)}$ given  $\mY_{(\mS_1)}$  and the last two terms  correspond to the marginal log-likelihood of $\mY_{(\mS_1)}$. Hence, $\wt Q(\mu, \bm \rho_{(\mS_{1})})$ can be seen as the full log-likelihood of $\mY$ and
we refer to the resulting estimator
as the full maximum likelihood estimator (FMLE). {\color{black} Note that if $|\mS_1|$ is   small enough, the contribution from the marginal distribution of  $\mY_{(\mS_1)}$ as seen in \eqref{ssar0} is small compared with that of $\mY_{(\mS_0)}$, and thus can be ignored.}  In this case, we will not lose too much by approximating $\wt Q(\mu, \bm \rho_{(\mS_{1})})$ by its first term as
\beq
\label{lsobj}
Q(\bm \rho_{(\mS_{1})})=\left\Vert \mY_{(\mS_0)}- A_{(\mS_0,\mS_{1})} \mD_{(\mS_{1})}  \mY_{(\mS_{1})} \right\Vert^2.
\eeq
This  is equivalent to optimizing $\bm \rho_{(\mS_{1})}$ conditional on
$\mY_{(\mS_1)}$. We will denote the resulting estimator as the conditional quasi-maximum likelihood estimator (CMLE).

Write  $ \mD_{(\mS_1)} \mY_{(\mS_{1})}  = \mbox{diag} ( \mY_{(\mS_{1})} ) \bm \rho_{(\mS_{1})}$ and denote the CMLE as $\hat{\bm\rho}\in\mR^{N}$ which has an explicit form in that  $\bm{\hat \rho}_{(\mS_0)}=0$ and
\beq
\label{cmleest}
\bm{ \hat \rho}_{(\mS_1)}= ( \bar \mX_{(\mS_{1})}^\top \bar\mX_{(\mS_{1})})^{-1} \bar\mX_{(\mS_{1})}^\top \mY_{(\mS_0)},
\eeq
where
$\bar\mX_{(\mS_{1})} = A_{(\mS_0,\mS_{1})} \mbox{diag} ( \mY_{(\mS_{1})} )\in\mR^{|\mS_0|\times |\mS_1|}$. As long as the number of the influential users is small which is intuitively true for many real-life networks, the CMLE   can be applied for very large networks.  In this paper, we assume
  $|\mS_1|=O(N^{\alpha_1})$ with $\alpha_1 \in [0,1)$  being a small constant.
 Based on Theorem \ref{thm:E.1}
 in Appendix \ref{sec:thmE.1}, we have
$\{\wt  Q(\mu, \bm \rho_{(\mS_{1})})-Q(\bm \rho_{(\mS_{1})})\}/Q(\bm \rho_{(\mS_{1})})=O_p(N^{\alpha_1-1}),$
which  requires that $\alpha_1<0.5$.

We know that  the difference between $Q(\bm \rho_{(\mS_{1})})$ and $\wt  Q(\bm \rho_{(\mS_{1})})$ is ignorable asymptotically, making the CMLE $\bm{\hat \rho}_{(\mS_1)}$ as efficient as FMLE.  The results  hold even when  $|\mS_1|$ tends to infinity at a slower rate than $N$.
 It also implies that for an arbitrary set $\mM$ satisfying  $\mS_1\subset \mM$,    endogeneity issues  can be ignored once writing
$\mY_{(\mM^c)}$ as a conditional model (i.e., model \eqref{work_model} below), and the conditional maximum likelihood estimation approach is still applicable. In the next section, we will use the conditional approach to estimate the unknown support $\mS_1$.


\subsection{Parameter estimation with unknown support}\label{unknown support estimation}
{\color{black}
We now discuss how to select a set of  candidate  users in the network which have nonzero $\rho$-coefficients. }
 Our method is based on the simple intuition that for a user to be influential, it should be well-connected. This motivates the use of a topology-based method for screening out potentially influential users in a first step.  Towards this,  set $\mM$ consist of users having  relatively large in-degree values (i.e., having large values in $|\mN_i|$ where $|\mN_i|=\sum_{j=1}^N a_{ji}$). By choosing a proper in-degree cut-off value, intuitively we will have  $\mS_1\subset \mM$. Formally this claim holds true under Condition (C3) in Appendix \ref{sec:condition} on
the minimum number of influential users. Operationally, we order the users according to their in-degree values $|\mN_i|$ and only retain those users whose values are among the largest $\lfloor N^\gamma \rfloor$.   Here $\gamma$ is  a tuning parameter, which will be chosen based on the network density and an in-degree histogram.   Theoretically, based on Conditions (C1)--(C6) in Appendix \ref{sec:condition}, a convenient choice is to set $\gamma \geq 5/9$,  and hence this screening step is conservative.
We discuss further in  Appendix \ref{sec:choice} the choice of $\gamma$ in real data analysis.

The identification of $\mM$ in the initial stage enables us to write  the  network influence regression model in \eqref{ssar1} as a conditional model
 \beqr
\mY_{(\mM^c)} =  A_{(\mM^c,\mM)} \mD_{(\mM)}  \mY_{(\mM)}  + \mE_{(\mM^c)}=\mX_{(\mM)} \bm \rho_{(\mM)}+\mE_{(\mM^c)}, \label{work_model}
\eeqr
where $\mM^c=\{1\leq i\leq N, i\notin \mM\}$ is the complement of  $\mM$ and $ \mX_{(\mM)}= A_{(\mM^c,\mM)} \mbox{diag} ( \mY_{(\mM)} )$ is the predictor matrix. Our goal is to trim down $\mM$ further to identify those elements of it  that are influential. Recognizing the similarity of the model above to linear regression, a natural idea is to employ a forward-addition algorithm for identifying the important variables in $\mM$, one at a time. Towards this, we introduce the following notations that will simplify presentation. For an arbitrary model $\mS$, we write $\RSS(\mS)=\mY_{(\mM^c)}^\top (\mI_{|\mM^c|}-\mH_{(\mS)})\mY_{(\mM^c)}$  as the residual sum of squares after fitting a linear regression model to $\{\mY_{(\mM^c)}, \mX_{(\mS)}\}$, where
  $\mH_{(\mS)}=\mX_{(\mS)} \{\mX_{(\mS)}^\top  \mX_{(\mS)}\}^{-1} \mX_{(\mS)}^\top$ with
$\mX_{(\mS)} = A_{(\mM^c,\mS)} \mbox{diag} ( \mY_{(\mS)} )\in\mR^{|\mM^c|\times |\mS|}$.
   The detailed algorithm is given in Algorithm \ref{alg:fr}.

\begin{algorithm}[!h]
\caption{The Forward-Addition Algorithm}
\label{alg:fr}
\begin{algorithmic}
\STATE 1. (Initialization). Set $\mS^{(0)}=\phi$.
\STATE 2. For $k = 1,2,\cdots,K$, repeat the following forward-addition iteration.
\begin{itemize}
\item [(2.a)]  Given $\mS^{(k-1)}$, for every $j\in \mM\backslash \mS^{(k-1)}$, construct a candidate model $\mJ^{(k)}_{j}=\mS^{(k-1)} \cup \{j\}$ and  obtain the residual sum of squares as $\mbox{RSS}(\mJ^{(k)}_{j})$;
\item [(2.b)]  Find $a_k=\argmin_{j\in \mM\backslash \mS^{(k-1)}}$ RSS$(\mJ^{(k)}_{j})$ ,  update
 $ \mS^{(k)}=\mJ^{(k)}_{a_k}$, and let $k \leftarrow k+1$.
\end{itemize}
\end{algorithmic}
\end{algorithm}

For a prespecified $K$, Algorithm \ref{alg:fr}  provides a solution path $\mS^{[K]}=\{\mS^{(k)},1 \leq k \leq K\}$, where $\mS^{(k)}=\{a_1, \cdots, a_k\}$. The candidate model is then selected from the solution path $\mS^{[K]}$. We have the following theorem that  shows the screening consistency of our forward-addition algorithm.
\begin{theorem} \label{theorem3.1}
 Under  Conditions (C1)--(C6) discussed in Appendix \ref{sec:condition},
 as $N \rightarrow \infty$, with probability tending to one, (1) the forward-addition algorithm finds a superset of $\mS_1$ in $m^*$ steps. That is,
	$P( \mS_1 \subset  \mS^{(m^*)}) \rightarrow 1$;
    (2) we have
	$P( \mS_1 \subset   \mS^{(k^*)}) \rightarrow 1$, where $k^*$  is the optimal forward-addition steps.
\end{theorem}
Here, $m^*=O(N^{2+5\alpha_1-2\alpha_2})$  is discussed  in Condition (C6) of Appendix \ref{sec:condition} and it represents   the minimal number of the forward steps in the forward-addition algorithm. 
From the discussion in  Appendix \ref{sec:condition}, we can take $m^*= O(N^{5/9})$ in theory.
In practice, a more conservative
$K$ (i.e., the number of  forward-addition steps) can be  chosen via the in-degree histogram for some $K \geq  \lfloor N^{5/9} \rfloor $ (i.e., $K\geq m^*$), where $\lfloor s \rfloor$ stands for the  largest integer no larger than $s$.
We employ a  BIC-type criteria to compare different  models  in $\mS^{[K]}$.
 Specifically,  motivated by \cite{chen2008EBIC},
 for a candidate  model $\mS$ under investigation,  we define a high dimensional extended BIC as
\beqr
\mbox{EBIC}(\mS)= \log\Big\{\RSS(\mS)\Big\}+|\mM^{c}|^{-1}|\mS|\Big\{\log(|\mM^c|)+2\log(|\mM|)\Big\}.
\n
\eeqr
Specifically,  $ k^*=\argmin_{1\leq k\leq K} \mbox{EBIC} (\mS^{(k)})$ is  the step that produces the smallest EBIC value.
 Theorem \ref{theorem3.1}(2)
 has established the  screening consistency result about $\mS^{(k^*)}$.

 Note that $\mS^{(k^*)}$ is the estimated model for $\mS_1$.
Theorem \ref{theorem3.1}(2) ensures that
 $\mS_1 \subset     \mS^{(k^*)} $   with probability approaching one. Although we are not able to show the selection consistency of the forward-addition algorithm in that the probability of  $\mS_1 = \mS^{(k^*)}$ approaches one,
 the numerical  results in Section \ref{sec:simulation} suggests that they can be quite close.  For an arbitrary model $\mS$, we denote
  $ \bar\mX_{(\mS)}= A_{(\mS^c,\mS)} \mbox{diag} ( \mY_{(\mS)} )$ as the predictor matrix.
After the forward-addition procedure,
we  have the following asymptotic normality about the estimated parameter $\hat{\bm\rho}_{(\mS^{(k^*)})} =\{{\bar \mX}_{(\mS^{(k^*)})}^\top{\bar \mX}_{(\mS^{(k^*)}) }\}^{-1}{\bar \mX}_{(\mS^{(k^*)})}^\top \mY_{(\mS^{(k^*)})}$.
\begin{corollary} \label{corollary 1}
 Under   Conditions (C1)--(C7) discussed in Appendix \ref{sec:condition},
$\hat{\bm\rho}_{(\mS^{(k^*)})} $  is asymptotically normal, in the sense that
$$|\mS^{(k^*)}|^{-1/2}B_N^{1/2}(\mS^{(k^*)})\wt\Psi \Lambda(\mS^{(k^*)})^{1/2}\mbox{diag}(\mY_{(\mS^{(k^*)})})\big(\hat {\bm\rho}_{(\mS^{(k^*)})}-\bm \rho_{(\mS^{(k^*)})}\big)\xrightarrow{d} N(0,\wt G) $$
 as $N \rightarrow \infty$, where $\wt\Psi$ is an arbitrary  $q \times |\mS^{(k^*)}|$  matrix with $q<\infty$ such that $|\mS^{(k^*)}|^{-1}\wt\Psi \wt\Psi^\top \rightarrow \wt G $ for some
 $q \times q $ nonnegative symmetric matrix $\wt G$, and $N(0,\wt G)$ is the multivariate normal distribution with mean $0$ and covariance matrix $\wt G$.
\end{corollary}
 Note that   $B_N(\mS^{(k^*)})=\diag(|\mN_j|, j\in\mS^{(k^*)})$ is a diagonal matrix, $\Lambda(\mS^{(k^*)})$ is  finite positive definite matrix and both are defined in Condition(C6) of  Appendix \ref{sec:condition}.
 The corollary  states that $\hat{\bm\rho}_{(\mS^{(k^*)})} $ as the CMLE estimator of ${\bm\rho}_{(\mS^{(k^*)})}$ follows asymptotically a normal distribution when $\mS_1$ is estimated as $\mS^{(k^*)}$.
 Conditional on the selected superset of influential
users,  the  standard error (SE)  and p-value   for each $\rho_i$ (the estimator of the
strength of influence) provided in Table \ref{tab:set_info} can be directly calculated using  Corollary \ref{corollary 1}.
This corollary is particularly valuable for providing disclaimers by quantifying the uncertainty associated with the strength of influence estimates.

\subsection{Comparison of Algorithm 1 with  other algorithms}

As discussed in  \cite{efron2004least},  there is a connection between
Algorithm 1 and Lasso. Specifically, a standard Lasso estimator solves  the following penalized optimization problem
\beqr
\Big\|\mY_{(\mM^c)} - \mX_{(\mM)} \bm \rho_{(\mM)}\Big\|^2+\lambda \big\|\bm \rho_{(\mM)}\big\|_1. \n
\eeqr
As the tuning parameter  $\lambda$  varies
from 0 to infinity, the solution path of Lasso sequentially covers a series
of models. The total number of models to be covered is typically of the same order as the number of variables  $p$ (i.e., $|\bm \rho_{(\mM)}|$) or when the degrees of freedom runs out. The best model is then selected from the solution
path. From this perspective, one key functionality of Lasso is to provide a solution
path, which is in spirit the same as the solution path provided by the Forward-Addition
 algorithm. Note that the computational complexity of the vanilla version of Lasso is
$O(Npm)$, where $m = O(\min\{N, p\})$ (see \cite{efron2004least}). This leads to a computational advantage  of Algorithm 1 as its  computational complexity is  $O(NpK)$, where $K$ is usually much smaller than $N$ and $p$.  In the case when early stopping is employed for Lasso  in order to choose a very sparse model, typically large values of the tuning parameter $\lambda$ will be employed. It is known that  the existence of the tuning parameter  $\lambda$  in a standard Lasso estimator  will introduce unnecessary bias.
These properties of the Lasso estimator have been well documented under various settings; see  \cite{zou2006the},  \cite{Wang2009},  and  \cite{guo}.
This is also evidenced numerically  by inferior  performance of  Lasso as compared with Algorithm 1,  which has been omitted to save space.

We have shown that our  Forward-Addition
 algorithm  can correctly identify  a superset of the true influential users. When this screening consistency holds,
 the  selection consistency property holds  only if  the true model is contained in the solution path. To ensure this, we can modify Algorithm 1 to allow backward elimination. For example, once a superset is identified according to EBIC, we can employ an elimination step for example by testing the adequacy of each variable in the model. However, this step is not needed in our numerical examples later on, since the extent to obtain over-fitted model is alreay quite small for our algorithm. We remark also that  Algorithm 1  is a gradient
based method which can be easily parallelized and thus is well suited for distributed computation. Intuitively for each step of the algorithm,  because only univariate regressions are needed to choose an optimal regressor and these regressions are independent of each other, each forward step can be computed in a distributed manner, allowing for efficient implementation and the application of our model to large scale social networks.
From the above considerations, we recommend Algorithm 1 in practice.

\section{Real Data Analysis} \label{sec:data}
In this section, we focus on the analysis of the Henan Floods data with its background introduced  in Section \ref{sec:data1} to identify influential users.
 The network is fixed and  we find that the  smallest, average    and largest in-degree value (i.e., $\mN_i$) are 0, 29, and 667 respectively, while on average,  each  user has 29 followers and is  followed by approximately   1.3\% users in the network.
 On the other hand, the smallest  and largest out-degree values (i.e., $\sum_{j}a_{ij}$) equal to  0 and 330, respectively.
 The response variables are the log number of reposts, comments, and likes, respectively and the total amount of Reposts, comments and likes are 3.6 $\times 10^7$, 1.3 $\times 10^7$, 5.3 $\times 10^7$, respectively.
We further check the empirical distributions of  responses. As can be seen from the histograms of the residuals and normal Q-Q plots in Fig \ref{fig4} in Appendix \ref{sec:subgaussion},  the response variables can be seen to follow some sub-Gaussian distributions.

To use our SNIR model,  we order the users according to their in-degree values  by taking $\gamma=2/3$ in the screening step, thus retaining  $\lfloor N^{2/3}\rfloor = 172$ users  to form the initial set $\mM$.
After that, we apply the forward-addition algorithm to the working model \eqref{work_model} to narrow down the influential user set.
{\color{black}  We first  compare  the SNIR model with the classical  SAR model
$ Y_i =  \rho \sum_{j = 1}^N w_{ij} Y_j + \varepsilon_i$  where
$ w_{ij}= a_{ij}/\sum_{j=1}^N a_{ij}$ and the scalar parameter $\rho$ can be seen as the average influence of neighbors.
It is found that the estimates of $\rho$ in SAR are all positive for different responses, suggesting that the posting
behaviors of connected users are positively related in the network. The results are consistent with the SNIR model results in that the estimated $\rho_i$'s in SNIR are always  positive across the responses. On the other hand, it is observed in Table  \ref{tab:R2} that the R$^2$ and adjusted R$^2$ for SNIR are always larger than the corresponding quantities for SAR, suggesting that our proposed influence may provide better fit to the data. We note that the rationale of using R$^2$ and adjusted R$^2$ is due to the similarity of these two models to the linear model. In Appendix C, we evaluate the impact of the choice of $\gamma$ in deciding the size of $\mM$ and find that the resulting influential users are more or less stable.
}

 \begin{table}[!htbp]
\bc\emph{}
\caption{\label{tab:R2}  Assessment of goodness-of-fit  for the sparse network influence regression (SNIR) model and the spatial autoregressive model (SAR) across three responses. Larger values correspond to a better fit.}
\vspace{0.15cm}
{\scriptsize
\begin{tabular}{cccccccccc}
\hline
  & \multicolumn{2}{c}{SNIR}&&  \multicolumn{2}{c}{SAR} \\
 Response &  R$^2$(\%)  &  Adjusted   R$^2$(\%) && R$^2$(\%)  &  Adjusted   R$^2$(\%)   \\
 \cline{2-3}\cline{5-6}
  \\
  \hline
  log number of Reposts  &               49.6  & 49.4 &&  21.6&  21.6 \\
  log number of  Comments  &           50.6  & 50.5&&   35.2&  35.2  \\
  log number of  Likes  &                50.7  & 50.5&&   29.4 & 29.4\\
\hline
\end{tabular}}
\ec
\end{table}

 The  identified influential users  for the SNIR model are listed in Table \ref{tab:set_info}.  Conditional on the selected superset of influential
users and Corollary \ref{corollary 1}, the SE of their influence parameter estimates together with their $p$-value are also presented.
The three estimated  influential user sets are denoted as  R, C and L with size  $|R|=10$, $|C|=9$ and $|L|=10$, respectively.
More details about the detected influential users are summarized in Table \ref{tab:user_info} below.

\begin{CJK*}{UTF8}{gbsn}
\begin{table}[!hbt]
\caption{Influential user sets with log-transformed reposts, comments and likes as  response, respectively. }
\begin{scriptsize}
\begin{tabular}{c|c|c|c|c|c|c}
\hline
Influential  Set  & User Name     & User Name In English & $\hat \rho_i$ & SE    & t value & p value   \\ \hline
 Set R  & TFBOYS-易烊千玺   & Qianxi Yiyang        & 0.111    & 0.021 & 5.237   & 1.782E-07 \\ \cline{2-7}
   (Reposts)              & 马伯庸           & Boyong Ma            & 0.368    & 0.051 & 7.210   & 7.582E-13 \\ \cline{2-7}
                 & 李荣浩           & Ronghao  Li          & 0.357    & 0.052 & 6.837   & 1.035E-11 \\ \cline{2-7}
                 & X玖少年团肖战DAYTOY & Zhan Xiao            & 0.144    & 0.026 & 5.479   & 4.749E-08 \\ \cline{2-7}
                 & BIGBIG张大大     & Dada Zhang           & 0.202    & 0.035 & 5.798   & 7.645E-09 \\ \cline{2-7}
                 & 何炅            & Jiong He             & 0.476    & 0.025 & 18.748  & 4.875E-73 \\ \cline{2-7}
                 & 韩国me2day      & me2day               & 0.362    & 0.074 & 4.898   & 1.036E-06 \\ \cline{2-7}
                 & 于正            & Zheng Yu             & 0.638    & 0.091 & 7.021   & 2.897E-12 \\ \cline{2-7}
                 & 杨迪            & Di Yang              & 0.279    & 0.064 & 4.394   & 1.164E-05 \\ \cline{2-7}
                 & 湖南卫视          & Hunan TV             & 0.434    & 0.073 & 5.939   & 3.308E-09 \\ \hline
Set C& THE9-虞书欣      & Shuxin Yu            & 0.167    & 0.037 & 4.471   & 8.168E-06 \\ \cline{2-7}
(Comments)                  & GAI周延         & Yan Zhou             & 0.169    & 0.045 & 3.778   & 1.622E-04 \\ \cline{2-7}
                 & 李荣浩           & Ronghao  Li          & 0.340    & 0.049 & 6.922   & 5.771E-12 \\ \cline{2-7}
                 & Alex是大叔       & Alex                 & 0.508    & 0.116 & 4.374   & 1.275E-05 \\ \cline{2-7}
                 & BIGBIG张大大     & Dada Zhang           & 0.212    & 0.039 & 5.416   & 6.736E-08 \\ \cline{2-7}
                 & 何炅            & Jiong He             & 0.527    & 0.026 & 20.338  & 1.276E-84 \\ \cline{2-7}
                 & 彭于晏           & Yuyan Peng           & 0.223    & 0.043 & 5.195   & 2.229E-07 \\ \cline{2-7}
                 & 于正            & Zheng Yu             & 0.618    & 0.071 & 8.692   & 6.688E-18 \\ \cline{2-7}
                 & 杨迪            & Di Yang              & 0.205    & 0.042 & 4.944   & 8.212E-07 \\ \hline
Set L & TFBOYS-易烊千玺   & Qianxi Yiyang        & 0.120    & 0.023 & 5.184   & 2.363E-07 \\ \cline{2-7}
(Likes)                  & GAI周延         & Yan Zhou             & 0.194    & 0.049 & 4.001   & 6.509E-05 \\ \cline{2-7}
                 & 李荣浩           & Ronghao  Li          & 0.286    & 0.044 & 6.503   & 9.649E-11 \\ \cline{2-7}
                 & Alex是大叔       & Alex                 & 0.382    & 0.088 & 4.347   & 1.441E-05 \\ \cline{2-7}
                 & BIGBIG张大大     & Dada Zhang           & 0.163    & 0.037 & 4.364   & 1.334E-05 \\ \cline{2-7}
                 & 何炅            & Jiong He             & 0.439    & 0.023 & 18.726  & 6.974E-73 \\ \cline{2-7}
                 & 彭于晏           & Yuyan Peng           & 0.167    & 0.040 & 4.137   & 3.647E-05 \\ \cline{2-7}
                 & 来去之间          & Gaofei Wang          & 0.368    & 0.061 & 6.070   & 1.494E-09 \\ \cline{2-7}
                 & 于正            & Zheng Yu             & 0.472    & 0.056 & 8.429   & 6.097E-17 \\ \cline{2-7}
                 & 杨迪            & Di Yang              & 0.195    & 0.040 & 4.922   & 9.181E-07 \\ \hline
\end{tabular}
\end{scriptsize}
\label{tab:set_info}
\end{table}
\end{CJK*}

We now include a simulation study based on real data to examine the relationship between the proportion of correctly detected influential users and the uncertainty surrounding their influence strength. Specifically, we use the response of reposts as an example. With $|R| = 10$, we fix the 10 selected influential users and randomly select one additional user from the candidate set $\mathcal{M}$ to act as a new influential user. The original estimators (i.e., $\hat{\rho}_i$s) serve as the coefficients for the selected influential users. For the new influential user, we evaluate 12 different coefficient values, generated as a sequence of integers from 1 to 12, each multiplied by 0.025. Using the original response values for the influential users, along with independently generated noise terms $\epsilon_i \sim N(0, 1)$, we simulate the responses of non-influential users. As the coefficient of the new influential user changes, the signal-to-noise ratio (SNR) of model \eqref{ssar1} also varies.

We repeat this process 100 times for each coefficient and calculate the proportion of correctly detecting the new influential user. The results, shown in Fig \ref{fig:proportion_snr} in Appendix \ref{sec:proportion}, are presented for the responses based on ``$Y_i = \log$ number of reposts, comments, and likes.''
As illustrated in the first subfigure of Fig \ref{fig:proportion_snr}, the proportion of correctly detected influential users increases with higher SNR values, demonstrating that our approach reliably identifies an influential user when its signal strength is sufficiently high. A similar trend is observed for responses based on comments and likes, as shown in the middle and right subfigures of Fig \ref{fig:proportion_snr} in Appendix \ref{sec:proportion}.

To validate the effectiveness of our approach for identifying influential users in this network, we compare its performance with several purely topology-based approaches and a purely response-based approach. For the topology-based approaches, we consider the classical in-degree method, betweenness centrality, and harmonic centrality for illustration. We can use standard R functions to calculate the betweenness and harmonic centrality for each user. For the in-degree method, we simply identify the users with the largest in-degree values as influential. Similarly, for betweenness and harmonic centrality, we select the users with the largest betweenness or harmonic values, respectively. For the response-based approach, we choose the users with the largest responses for each response variable. To ensure comparability with the SNIR method, the number of users declared as influential by these five approaches is set to match the size of the corresponding set identified by our approach (i.e., $|R|$, $|C|$, and $|L|$). For example, for the response variable "comments," our method identifies 9 influential users. The topology-based methods then simply select the 9 users with the largest in-degree, betweenness, or harmonic values as influential, and the response-based method chooses the 9 users with the most comments.

Since the influential users identified by topology-based and response-based methods are not selected through a probabilistic model, the associated standard errors of their influence parameter estimates are not known.
To assess the performance of each approach more comprehensively, we evaluate them from two additional perspectives for each response.
 The first measures the loss of followers if the identified influential users are removed from the network.
The second measure is the change of the response if the influential users are removed, taking into account of the loss of the response due to the removal of the influential users themselves, as well as the influence that these users bring to the network.  For the latter, we  use  CMLE to regress away the influence of the influential users on non-influential ones, using the sets identified by each method.  Apparently, for both measurements, the larger they are, the better an approach is. The results are summarized in  Table \ref{tab:network_effect} below.  To save space, we use log(Reposts)/log(Comments)/log(Likes) to represent $Y_i$, the log number of reposts/comments/likes.
We observe that the removal of the influential users identified by the SNIR method has the largest impact on network information spreading in terms of the amount of response lost. The SNIR method has a weaker impact on the number of followers compared to the topology-based methods, which is reasonable as the topology-based methods tend to select users with the largest in-degree values.
We also see that, for this network, the topology-based methods outperform the response-based approach in most cases. The SNIR method outperforms the response-based approach for all the response variables and measurements. This suggests that the structure of the network may play a bigger role in determining nodal influence.

\begin{CJK*}{UTF8}{gbsn}
\begin{table}[!h]
\caption{Detailed information for the identified influential users. Followers: the number of followers, In-degree: the in-degree value, Likes: the number of likes, Comments: the number of Comments, Reposts: the number of reposts, Membership: R for reposts, C for comments and L for likes. }
\centering
\begin{scriptsize}
\begin{tabular}{c|c|c|c|c|c|c|c}
\hline
User Name     & In English & Followers    & In-degree  & Likes   & Comments & Reposts & Membership             \\ \hline
BIGBIG张大大     & Dada Zhang           & 32441000  & 179  & 145086  & 9119     & 93294   & R, C, L \\ \hline
何炅            & Jiong He             & 120000000 & 667  & 142891  & 5327     & 15418   & R, C, L\\ \hline
李荣浩           & Ronghao Li           & 23537109  & 205  & 10613   & 760      & 1065    & R, C, L \\ \hline
杨迪            & Di Yang              & 7973000   & 223   & 27324   & 2560     & 317     & R, C, L \\ \hline
于正            & Zheng Yu             & 8176000   & 236    & 872     & 72       & 42      & R, C, L \\ \hline
TFBOYS-易烊千玺   & Qianxi Yiyang        & 88623341  & 349  & 1397934 & 125832   & 1109797 & R, L        \\ \hline
彭于晏           & Yuyan Peng           & 32944000  & 242   & 13601   & 1256     & 1245    & C, L         \\ \hline
Alex是大叔       & Alex                 & 17873113  & 104   & 483     & 40       & 387     & C, L       \\ \hline
GAI周延         & Yan Zhou             & 11379901  & 106    & 67597   & 16550    & 26437   & C,  L        \\ \hline
X玖少年团肖战DAYTOY & Zhan Xiao        & 29547091  & 201  & 5389162 & 1000000  & 1000000 & R               \\ \hline
韩国me2day      & me2day               & 26204000  & 145   & 1155    & 56       & 310     & R               \\ \hline
湖南卫视          & Hunan TV             & 15996639 & 210  & 823     & 437      & 156     & R               \\ \hline
马伯庸           & Boyong Ma            & 7779218   & 184 & 607     & 51       & 1262    & R              \\ \hline
THE9-虞书欣      & Shuxin Yu            & 15724775  & 138 & 78847   & 26752    & 13861   & C               \\ \hline
来去之间          & Gaofei Wang          & 768000   & 243  & 339     & 73       & 1564    & L              \\ \hline
\end{tabular}
\end{scriptsize}

\label{tab:user_info}
\end{table}
\end{CJK*}

\begin{table}[!h]
\renewcommand\arraystretch{1.5}
\bc\emph{}
\caption{\label{tab:network_effect} A comparison of  five approaches for identifying influential users in the Henan Floods network:  $\Delta$(R) represents the percentage of the responses lost due to the removal of influential users, while $\Delta$(F) represents the percentage of followers lost.}
\vspace{0.25cm}
{\scriptsize
\begin{tabular}{ccccccccccccccc}
\hline
 Response &  \multicolumn{2}{c}{SNIR}&&  \multicolumn{2}{c}{In degree-based} &&  \multicolumn{2}{c}{Response-based}&&  \multicolumn{2}{c}{Betweenness-based}&&  \multicolumn{2}{c}{Harmonic-based}\\
& $\Delta$(R) & $\Delta$(F) &&$\Delta$(R) & $\Delta$(F) &&$\Delta$(R) & $\Delta$(F) &&$\Delta$(R) & $\Delta$(F) &&$\Delta$(R) & $\Delta$(F) \\
 \cline{2-3}\cline{5-6}\cline{8-9}
 \cline{11-12}\cline{14-15}
  \\
  \hline
  log(Reposts)  &           76.0\%&      5.6\%&&     64.6\%&      7.0\%&&     65.2\%&      2.9\%&&     68.1\%&      6.0\%&&         63.4\%&      5.1\%\\
log(Comments) &       88.0\%&      4.8\%&&     81.4\%&      6.6\%&&     71.4\%&      2.8\%&&     84.1\%&      5.5\%&&         81.4\%&      4.7\%\\
log(Likes)&                   89.2\%&      5.8\%&&     80.9\%&      7.0\%&&     74.0\%&      3.2\%&&     82.6\%&      6.0\%&&         79.6\%&      5.1\%\\
\hline
\end{tabular}}
\ec
\end{table}
We zoom in to examine the selected influential users while reposts is used as the response  (i.e., the  set R).  We find  that they represent a variety of popular figures and celebrities in different fields, including  pop stars, a senior overseas entertainment account, a senior director,   CEOs, a well-known novelist, and an official account of a popular TV channel in China.
If we use the response-based  selection approach, two pop stars (Qianxi Yiyang and Zhan Xiao) in set R will be selected as influential users since they have the largest number of reposts. Incidentally,  the
commercial values of these two are among  the top 5   celebrities in 2020-2021  as reported by  Aiman  Data Survey, an official   survey company officially associated with Sina Weibo.
 On the other hand, if we directly use the  topology-based approach based on in-degree values to select influential users and retain the top 10 users with the largest in-degrees,
two users (Jiong He and Qianxi
Yiyang) in  set R will also  be selected.   Thus, SNIR method can identify certain users that are identified as influential using either a topology-based or a response-based method.
On the other hand, some of the influential users identified by the SNIR method were not included in the lists of the top ten users with either the largest number of followers (as identified by the topology-based method) or the largest number of reposts (as identified by the response-based method). Notable examples include Gaofei Wang, the CEO of Sina Weibo, and Zheng Yu, the CEO of an entertainment company managing a large number of pop stars. Other influential users identified by the SNIR method, but not by the topology-based or response-based methods, include Di Yang and Ronghao Li, whose posts were liked, commented on, or reposted by other social influencers, contributing to a wider dissemination.

Finally, this dataset contains richer information because the timestamps of the responses are available. An obvious extension of the model is to investigate whether and how this dynamic information may impact the selection of influential users. In Appendix \ref{sec:dynamic model}, we discuss this extension and find conclusions qualitatively similar to those presented in this subsection. We also note that our modelling framework is extremely flexible when it comes to incorporating covariate information and flexible error structures. We illustrate another extension of the model in this direction in Appendix \ref{sec:withX}.

\section{Simulation} \label{sec:simulation}

We conduct extensive simulation to study the performance of our SNIR model for identifying
influential users. We do it from two perspectives. The first is via generating synthetic data and
the second is to simulate data following the network structure of the Henan flood data.

\subsection{Synthetic data}\label{sec:simulation1}
For the simulation in this section, we  present examples by generating data from
the model in \eqref{ssar}.
For each example, we set $|\mS_1| = 10$ or $15$. For screening, we set  $|\mM|=\lfloor N^{\gamma}\rfloor $ with $\gamma=2/3$; that is, we choose $|\mM|$ users with the largest in-degree values as the initial set of influential users before the forward-addition algorithm is employed. Note that in the simulation below $N$ is set to be large and thus $|\mM|$ is much larger than $|\mS_1|$.
The true influential users collected in $\mS_1$ are chosen uniformly at random from $\mM$.
For each   $ i \in \mS_1$,  the influential coefficient  $\rho_{i}$  is generated from a uniform distribution $U(0.5,1)$. The error term $\ve_i$ is independently  generated from $N(0,1)$.
 We choose $\mu=5$ in the model to ensure that  the response values for all users  are not too close to zero, motivated by our real dataset.
For the network,   we consider the following three structures.

{\sc Example 1.} (ER Model). We start with the simplest Erd\H{o}s-R\'enyi model \citep{Erdos:Renyi:1959}, where edges are  independently generated with probability $P(a_{ij}=1)=0.5 N^{-0.8}$.   We consider three network sizes with $N=5000, 7500$  or $15000$.
As the network size increases from $N=5000$ to $N=15000$,  the network density decreases from 7.5\%  to  2.9\%.

{\sc Example 2.} (Stochastic Block Model).
The stochastic block model is useful when users in a network are clustered, and is one of the most popular network models  \citep{Nowicki:Snijders:2001}. We simulate data from a stochastic block model with $5$  blocks by dividing all the users uniformly at random to one of the blocks. The probability for two users to make a connection  is set as $P(a_{ij}=1)=N^{-0.8}$ if they are from the same block, and as $P(a_{ij}=1)=0.5N^{-0.8}$ otherwise. We consider $N=1000, 2500$ or $5000$.
As the network size increases from $N=1000$ to $N=5000$,  the network density  decreases  from 45\%  to  12\%.

{\sc Example 3.} (Power-Law Distribution). Many real-life networks have their degree values roughly following a power-law  distribution \citep{adamic2000power} and this example is to assess the performance of our approach when the network follows this structure. Towards this,  we first generate the in-degree value $|\mN_i| = \sum_j a_{ji}$ for user $i$ from a discrete power-law
distribution with $P(|\mN_i|  = k) = ck^{-\alpha}$, where $c$ is a normalizing constant and the exponent
parameter is set as $\alpha = 2.5$. Then, for the $i$th user, we choose $|\mN_i| $ users uniformly at random as its followers.
 We consider three network sizes as $N=5000, 10000$ or $20000$.
 As the network size increases from $N=5000$ to $N=20000$,  the network density  changes  from 7.8\%  to  2.0\%.

For each simulation setup,  100 datasets are generated and the results are averaged. We run the simulation on a PC with 3.70 GHZ CPU and 64G RAM. Because efficient computation is paramount to deploy SNIR method  for analyzing huge social networks, we record the average running time  and find that  on average, it takes less than one second for the analysis of a simulated dataset in all scenarios.

Denote   $\hat \mS_1= \mS^{(k^*)}$ as the estimated support of the influential user set chosen  by EBIC  via our forward-addition algorithm.
For performance measures, we record
(i) the true positive rate (TPR) defined as the proportion of the influential users correctly identified as influential (i.e., TPR$= |\hat \mS_1\cap \mS_1 |/|\mS_1|$); (ii)  the false positive rate (FPR)  defined as the proportion of the non-influential users incorrectly identified as influential (i.e., FPR$= |\hat \mS_1\cap (\mM \backslash {\mS_1}) |/|\mM \backslash {\mS_1}|$),
(iii) the correctly fitting probability (CFP) as the proportion of the simulation replicates where $\mS_1$ is exactly  identified  (i.e., CFP$=  \mathbf{I} (\hat \mS_1=\mS_1)$); and  (iv) the error (Err) in estimating $\bm\rho$ defined as $\| \hat{\bm \rho}- \bm \rho\|$.

\begin{table}[!h]
\renewcommand\arraystretch{1.5}
\bc\emph{}
\caption{\label{tab:t0} Simulation results.}
\vspace{0.25cm}
{\scriptsize
\begin{tabular}{ccccHHcccccHHc}
\hline
$N$& TPR & FPR&CFP& OF& UF&Err& &TPR& FPR&CFP&OF& UF& Err  \\
  \hline
 \multicolumn{14}{c}{Example 1: ER model}\\
& \multicolumn{6}{c}{$|\mS_1|=10$} && \multicolumn{6}{c}{$|\mS_1|=15$} \\\cline{2-7}\cline{9-14}
 5000 & 0.954& 0.001& 0.530&1.000&1.000& 0.043&&0.966 &  0.000 &      0.490&0.000 & 1.000 & 0.510   \\
 7500 & 0.987& 0.000& 0.870&0.000&1.000& 0.025&& 0.989 & 0.000 &      0.840& 0.000 &1.000 &  0.038   \\
 15000 & 1.000& 0.000& 1.000&0.000&0.000& 0.016&&1.000 & 0.001 &
 0.990 & 1.000& 0.000& 0.025 \\
 \hline
 \multicolumn{14}{c}{Example 2: stochastic block model}\\
 & \multicolumn{6}{c}{$|\mS_1|=10$} && \multicolumn{6}{c}{$|\mS_1|=15$} \\\cline{2-7}\cline{9-14}
 1000 & 0.964  & 0.000 &  0.640 &0.000 & 1.000 &   0.081 &&
        0.984 &  0.000 & 0.760 & 0.000 &  1.000& 0.107  \\
 2500 &1.000  & 0.001 &   0.990 &1.000 & 0.000 & 0.040 &&
       1.000  & 0.001 &   0.990 &1.000 & 0.000 & 0.061\\
 5000& 1.000  & 0.000 &  1.000  &0.000 & 0.000 & 0.028&&
        1.000 & 0.000 &  1.000  &0.000 & 0.000 & 0.043 \\
 \hline
\multicolumn{14}{c}{Example 3: power-law distribution network}\\
& \multicolumn{6}{c}{$|\mS_1|=10$} && \multicolumn{6}{c}{$|\mS_1|=15$} \\\cline{2-7}\cline{9-14}
5000 &0.957 &0.000 & 0.570 & 0.000 & 1.000 &0.042 &
     & 0.977&0.000 & 0.650 & 0.000 & 1.000 & 0.049\\
10000&0.988&0.000  & 0.880 & 0.000 & 1.000 &0.022 &
     &0.993& 0.000 & 0.890 & 0.000 & 1.000 &0.032 \\
20000&0.998&0.000  & 0.980 & 0.000 & 1.000 &0.014 &
     &0.999& 0.000 & 0.990 & 1.000 & 0.000 &0.022\\
 \hline
\end{tabular}}
\ec
\end{table}

From the results summarized in Table \ref{tab:t0},
the following conclusions can be drawn.
First, the average values of TPR  are always close to 100\%,  suggesting that SNIR method can identify the majority of the influential users.
On the other hand,  the average  FPR is 0\% or very close to it, indicating that non-influential users are hardly identified as influential.
 The proportions of the support of $\bm \rho$ being exactly identified increases as the size of a network increases, while the error of its estimation decreases. Thus, we conclude that SNIR method performs well in these simulated examples.

\subsection{Simulation based  on the real data}\label{simu:real}
Motivated by the real data analysis, we generate data using the network of the Henan floods dataset.
We consider three different settings to generate $\mS_1$ with $|\mS_1|=8$ as follows.

 {\bf Setting 1}.  We rank the response values of  the $N$ users and  take $|\mS_1|$ users with the largest response values as influential users.

{\bf Setting 2}.
 We pick the  $|\mS_1|$  users  with the largest number of followers in the network as influential users.

 {\bf Setting 3}. We
 randomly pick $|\mS_1|$ users as influential users from the candidate set $\mM$, where
 $|\mM|=\lfloor N^{2/3} \rfloor$.

For the responses, if $i\in\mS_1$, we use  the  number of log-transformed reposts/comments/likes in the real data as the response value.
For other response values (i.e.,  $Y_i$ with $i\in\mS_0$), we generate them according to  model  \eqref{ssar1}, where  $A$ is taken as the network of the Henan flood dataset.  Note that for Setting 1, to make sure that the users with the largest responses are indeed the influential users as in our model, we change the simulation slightly by generating  2 nonzero $\rho_i$s independently   from a uniform distribution $U(0.25,0.5)$ and the other $|\mS_1|-2$  nonzero $\rho_i$s independently   from  $U(-1,-0.5)$.
The random noises  $\ve_i$s are generated from $N(0,1)$.
 Setting 1 works the best for response-based approaches that rank influentialness based on the magnitude of the response, while Setting 2 favours the network-based approach that selects users based on connections. These two settings are used to assess the performance of SNIR  when they are not in favour of the proposed method.   We generate 100 random datasets this way and summarize the results in Table \ref{tab:real_network}.

\begin{table}[!h]
\renewcommand\arraystretch{1.5}
\bc\emph{}
\caption{\label{tab:real_network} Detailed results for Henan Flood network structure with  $|\mS_1|=8$. }
\vspace{0.25cm}
{\scriptsize
\begin{tabular}{ccccccccccccccc}
\hline
Response& TPR & FPR&CFP& Err& &TPR& FPR&CFP& Err& &TPR& FPR&CFP& Err  \\
  \hline
 \multicolumn{15}{c}{Setting 1(favouring response-based method)}\\
& \multicolumn{4}{c}{SNIR} && \multicolumn{4}{c}{Response-based}&& \multicolumn{4}{c}{Topology-based} \\
\cline{2-5}\cline{7-10}\cline{12-15}
Reposts &1.000 &0.000 &1.000 &0.011 &&1.000 &  0.000 &      1.000 & 0.011&&0.389 &  0.611 &      0.000 &      1.767   \\
Comments&1.000 &0.000 &1.000 &0.012 &&1.000 &  0.000 &      1.000 & 0.012&&0.540 &  0.460 &      0.000 &      1.729   \\
Likes &  1.000 &0.001 &0.990 &0.011 &&1.000 &  0.000 &      1.000 & 0.011&&0.529 &  0.471 &      0.000 &      1.544   \\
 \hline
  \multicolumn{15}{c}{Setting 2(favouring topology-based method)}\\
& \multicolumn{4}{c}{SNIR} && \multicolumn{4}{c}{Response-based}&& \multicolumn{4}{c}{Topology-based} \\
\cline{2-5}\cline{7-10}\cline{12-15}
Reposts  &1.000 &0.001 & 0.990&0.015 &&0.858 &  0.142 &      0.000 & 0.820&&1.000 &  0.000 &      1.000 &  0.015   \\
Comments &1.000 &0.001 & 0.990&0.016 &&0.949 &  0.051 &      0.610 & 0.019&&1.000 &  0.000 &      1.000 &  0.016   \\
Likes &   1.000 &0.001 & 0.990&0.013 &&0.969 &  0.031 &      0.750 & 0.015&&1.000 &  0.000 &      1.000 &  0.013   \\
 \hline
  \multicolumn{15}{c}{Setting 3(favouring the SNIR method)}\\
& \multicolumn{4}{c}{SNIR} && \multicolumn{4}{c}{Response-based}&& \multicolumn{4}{c}{Topology-based} \\
\cline{2-5}\cline{7-10}\cline{12-15}
Reposts & 1.000 & 0.001& 0.990 &0.018 &&0.892 &  0.107 &      0.300 &  0.275&&0.404 &  0.596 &      0.000 &      1.723   \\
Comments& 0.998 & 0.001& 0.970 &0.020 &&0.877 &  0.122 &      0.270 &  0.397&&0.404 &  0.596 &      0.000 &      1.714   \\
Likes &   1.000 & 0.001& 0.990 &0.015 &&0.940 &  0.060 &      0.570 &  0.021&&0.404 &  0.596 &      0.000 &      1.709 \\
 \hline
\end{tabular}}
\ec
\end{table}

  From this table, we can see that our  proposed SNIR model can identify the set of  influential users successfully in all the circumstances, with all the
TPRs equal to (or very close to ) 100\% and FPRs equal to 0\% almost.   As expected,  the  response-based and  topology-based methods perform
well under  a setting in their favour, but fail to capture many influential users otherwise.
    On the other hand,  
 our SNIR model is always  competitive with  the best method even under  Setting 1 or 2.
   These together illustrate the reliability of our SNIR model  to select truly significant users under different settings.

\section{Discussion} \label{sec:discussion}

In this paper,  we have proposed a novel  SNIR model to identify the   influential users in  a social network to capture task-specific influentialness, leveraging user connectiveness and nodal responses.
 To efficiently estimate the parameters in SNIR model, we have devised a highly computationally efficient procedure that can be easily scaled to large networks. We have successfully applied our method to a Sina Weibo data and found interesting phenomena.   


There are a few avenues for future research.
First, the model can be substantially generalized. One extension is to include covariates for which we provide additional simulation results in the supplementary material. 
 Second,  our use of the SNIR model for identifying three sets of influential users separately in the real data analysis can be potentially improved, possibly by stacking the three responses into a matrix. In this case, we may be able to exploit their correlation by developing a SNIR model for matrix responses.
Last but not the least, when a dataset is extremely large, the need to divide computation possibly in a distributed fashion on different machines arises. How to develop a computationally efficient procedure exploiting distributed computing remains an important question to explore.

{\bf Acknowledgements}: We thank Prof. Thomas Brendan Murphy, the editor, an associate editor, and two referees for their constructive comments that have greatly improved this paper.
\begin{appendix}
\section*{}\label{appn} 
The appendix is divided into several parts. Appendix \ref{sec:condition} contains theoretical conditions, Appendix \ref{sec:dynamic model} includes an extension of the SNIR model to handle dynamic data, and Appendix \ref{sec:choice} discusses the choice of the candidate set $\mM$. Appendix \ref{sec:withX} extends the model to include covariates and heterogeneous random errors. Appendices \ref{sec:thmE.1} through \ref{sec:corollary} contain proofs of two additional theorems, Theorem \ref{theorem3.1}, and Corollary \ref{corollary 1}.  Appendix \ref{sec:subgaussion}  includes an assessment of the Gaussianity of the data, while Appendix \ref{sec:proportion} presents additional simulation results based on real data.
To simplify notation, let $I$ represent the identity matrix of the appropriate dimension in the remaining proof.
\end{appendix}
\begin{appendix}

\section{Theoretical Conditions} \label{sec:condition}
In order to prove Theorems 3.1--3.3 and  Corollary 1, we impose the following conditions. 

\begin{itemize}

	\item[(C1)] The $\varepsilon_i$'s are i.i.d. sub-Gaussian random variables. That is, there exist two finite positive constants $\gamma$ and $C_1$ such that $E\{\exp(\gamma\epsilon_i^2)\}<C_1$ for all $i$.

\item[(C2)] There exist finite positive constants $\lambda_a$
and $\lambda_b$ such that $0<\lambda_a < \lambda_{\min}(\Sigma)\leq \lambda_{\max}(\Sigma)<\lambda_b<\infty$, where $\Sigma=\mbox{cov}(\mY)$.

	\item [(C3)]  As $N \rightarrow \infty $,  (i) $|\mS_1| = O(N^{\alpha_1})$ for some $\alpha_1<1$;
and (ii) there exists a finite positive constant $C_2$ such that $\mN_{\min}=\min_{i\in \mS_1} |\mN_i|\geq N/C_2$.

\item [(C4)] $N^{-1}A_{(\mS_{0},\mS_{1})}^{\top} A_{(\mS_{0},\mS_{1})}$
converges to a finite positive definite matrix $\Lambda_1$ in the Frobenius norm, that is,
$\|N^{-1}A_{(\mS_{0},\mS_{1})}^{\top} A_{(\mS_{0},\mS_{1})}-\Lambda_1\|_F=o(1)$, where
$0<\tau_{\min,1}<\lambda_{\min}(\Lambda_1)\leq \lambda_{\max}(\Lambda_1)<\tau_{\max,1}<\infty$
for some finite constants $\tau_{\max,1}$ and $\tau_{\min,1}$.

\item [(C5)] For $\bm\rho$, $1>\rho_{\max}\geq \rho_{\min}>  0$, where $\rho_{\min}=\min_{i\in\mS_1} |\rho_{i}|$ and $\rho_{\max}=\max_{i\in\mS_1} |\rho_{i}|$
     are finite positive constants.

\item [(C6)] Write $B_N(\mS)=\diag(|\mN_j|, j\in\mS)$ for any $\mS\subset \mM$, where $\mM$ is the candidate set defined in  \eqref{work_model}.
Assume
(i) there exists a finite positive constant $C_3$ such that $\min_{j\in\mM} |\mN_j|\geq N^{\alpha_2}/C_3$ for some $0<\alpha_2\leq 1$;
(ii) $B^{-1}_N(\mS) A_{(\mM^c, \mS)}^\top A_{(\mM^c, \mS)}$
converges to a positive matrix $\Lambda(\mS)$ uniformly in the Frobenius norm for any $\mS$ satisfying $|\mS|\leq m^*$, that is,
$\sup_{|\mS|\leq m^*} \|B^{-1}_N(\mS) A_{(\mM^c, \mS)}^\top A_{(\mM^c, \mS)}-\Lambda(\mS)\|_F=o(1)$, where
$\min_{|\mS|\leq m^*}\lambda_{\min}\{\Lambda(\mS)\}>\tau_{\min,2}$
for finite positive constant
$\tau_{\min,2}$ and $m^*=O(N^{m})$  represents the minimal number of steps required by the forward-addition algorithm with $m=2+5\alpha_1-2\alpha_2$ and $9\alpha_1+4<5\alpha_2$.

\item[(C7)]
Define $\wt\Omega=B_N^{-1/2}(\mS)\Lambda^{-1/2}(\mS) A_{(\mS^c,\mS)}^{\top}=(\wt\Omega_{ij})\in\mR^{|\mS|\times |\mS^c|}$ for a generic
set $\mS \subseteq\mS^{[m^*]}$.
There exists a finite positive constant $\alpha_3$ satisfying
$\max_{1\leq i\leq |\mS|} \sum_j \wt\Omega_{ij}^4=O(N^{-\alpha_3})$ with
$2m<\alpha_3\leq 1$ uniformly for  $\mS\supseteq\mS_1$.
\end{itemize}


Condition (C1) is weaker than the commonly assumed Gaussian
condition. Similar condition has also been used in \cite{yu2012}, \cite{su2012sieve},  and \cite{dou2016generalized}.
Condition (C2) assures the covariance matrix of $\mY$ has bounded eigenvalues. This condition is commonly assumed in existing  literature;
see, e.g., \cite{lee2004asymptotic},  \cite{yu2008quasi} and \cite{guo}.
 Condition (C3)(i) allows the number of true influential users tends to infinity
but at a  rate smaller than $N$. Condition (C3)(ii) requires that the minimum  number of followers of an influential user should be of the same order as the total network size $N$; that is, an influential user should have  sufficient
number of followers.

Condition (C4) is a technical  condition, otherwise the influential users in SNIR model cannot be identified.
 This condition plays a similar role for controlling the eigenvalues of covariate matrices in traditional linear regression models;
 see, e.g., \cite{fan2001variable} and \cite{zou2006the}.
 Condition (C5) assumes that all $\rho_i$s are within $(-1,1)$. Otherwise, the resulting model will not be stationary.  This type of condition is commonly assumed in existing literature; see, e.g,
\cite{yu2012}, \cite{dou2016generalized} and \cite{lewbel2021social}.

Condition (C6) is a technical condition and parallel to Condition (C4)(i).
This condition plays a similar role as the Sparse Riesz Condition  defined in \cite{zhang2008the}  for linear regression.
It is crucial for proving the screening consistency of the proposed forward-addition procedure.
Condition (C6)(i) requires that the users in the candidate set $\mM$  should have sufficient  number of followers. Condition (C6)(ii) is a standard requirement for  the forward-addition procedure,  otherwise, it can not be proceed successfully.  Similar conditions can be found in \cite{Wang2009}, \cite{zhang2009adaptive},
\cite{gao2019banded} and \cite{ma2021sparse}. It imposes some constrict on the orders of $\alpha_1$ and $\alpha_2$,
which is satisfied as long as $\alpha_1$ is small and $\alpha_2$ is large (e.g.,
$\alpha_1=0.1$ and $\alpha_2=1$).
 Condition (C7)  is used to establish the asymptotic normality of $\widehat{\bm\rho}_{(\mS^{(k^*)})}$.
 By Condition (C7), we have $\wt\Omega\wt\Omega^\top\rightarrow \mI_{|\mS|}$ such that $\sum_j \wt\Omega_{ij}^2\rightarrow 1$ uniformly for any $i$.  Condition (C7) can be satisfied as long as $\wt\Omega_{ij}$'s
 are of the same order with $\wt\Omega_{ij}=O(N^{-1/2})$ and $\alpha_3=1$.  Intuitively, for a sufficient  sparse network, Condition (C7) can be naturally satisfied.

All the above conditions can also be verified by the real dataset. For example, let us examine Condition (C4), (C6) and (C7) by
selecting a sub-network $\wt A$ form $A$ with size $N=500,1000$ or $2000$ after choosing the users uniformly at random,  where $A$ is the network structure defined in Section \ref{sec:data1}.
 Repeating the process for 100 times,  we find that the minimum and maximum eigenvalues of $N^{-1} \wt A_{(\mS_{0},\mS_{1})}^{\top} \wt A_{(\mS_{0},\mS_{1})}$
 are consistently around $0.050$ and $0.420$, respectively. This implies that  a positive definite $\Lambda_1$ exists such $N^{-1}A_{(\mS_{0},\mS_{1})}^{\top} A_{(\mS_{0},\mS_{1})}$ can be approximated by it.
For Condition (C6), we can directly set $\alpha_2=1$ in the real dataset and  verify that
 $\min_{i\in\mM} |\mN_i|/N$ is always close to 1/2 for different sub-network size  with 100 random replications. Accordingly, it implies that
  $\min_{i\in\mM} |\mN_i|/N$  can also be bounded away from $1/C_3$.  Similarly, we verify that the minimum and maximum eigenvalue of  $B^{-1}_N(\mS) \wt A_{(\mM^c, \mS)}^\top \wt A_{(\mM^c, \mS)}$
  are always very close to $0.010$  and
  $2$, respectively.
  Hence,  $\Lambda (\mS)$
 exists and Condition (C7) can be empirically  verified.

\section{A Dynamic Model}
\label{sec:dynamic model}
In the Henan floods dataset, as in many other datasets of a similar nature, there is rich information such as the posting times of users. To incorporate this,  we investigate an extension of our model to the dynamic scenario.  Based on the posting time information,  we have decided to split the entire dataset into two parts, denoted as $\mJ_1$  and $\mJ_2$ respectively, such that the number of events in each is about the same. This consideration leads to a sub dataset consisting of the data on the first day of the events with  $|\mJ_1|=1146$ and the second with $|\mJ_2|=1129$.  Denote $\mJ=\mJ_1 \cup \mJ_2$. Define $\mM=\mM_1\cup \mM_2 $ as the associated initial screening sets. Recall $\mM$ is obtained by  ordering  the users according to their in-degree values before retaining   $\lfloor N^{2/3}\rfloor = 172$ users with the largest in-degree values. We choose $\mM_1$ and $\mM_2 $ as the corresponding subsets of $\mM$  based on  the posting time.
 To  capture the dynamic information, we modify  the SNIR model as  follows
 \beqr
 \mY_{(\wt \mJ_1)} &=&  A_{(\wt \mJ_1,\mM_1)} \mD_{(\mM_1)}  \mY_{(\mM_1)}  + \mE_{(\wt\mJ_1)},\label{dyn:model1}\\
\mY_{(\wt \mJ_2)} &=&   A_{(\wt \mJ_2,\mM_2)} \wt \mD_{(\mM_2)}  \mY_{(\mM_2)}  + A_{(\wt \mJ_2,\mM_1)} \wt \mD_{(\mM_1)}  \mY_{(\mM_1)}  + \mE_{(\wt \mJ_2)},
\label{dyn:model2}
\eeqr
where  $\wt \mJ_1=\mJ_1\backslash \mM_1 $, $\wt \mJ_2=\mJ_2\backslash \mM_2 $, $\mD_{(\mM_1)}=\diag(d_j,j\in \mM_1)  $, $\wt \mD_{(\mM_1)}=\diag(\wt d_j,j\in \mM_1) $,  and $\wt \mD_{(\mM_2)}=\diag(\wt d_j,j\in \mM_2) $ are diagonal matrices with unknown coefficients $\bm d=(d_1,\cdots,d_N)^\top$ and $\wt {\bm d}=(\wt d_1, \cdots,\wt d_N)^\top$.  We set $d_j=0$ if $j\notin \mM_1$ and $\wt d_j=0$ if $j\notin \mM_1\cup \mM_2$.
In the formulation in \eqref{dyn:model2}, we have allowed those users exerting influence in $\mM_1$ to continue to exert influence in the second period.

We apply the forward-addition algorithm to the above working models \eqref{dyn:model1}--\eqref{dyn:model2} to identify the influential user set. The results are summarized in Table \ref{dynamic:user_set_info} and \ref{dynamic:network_effect}. Compared with Table \ref{tab:user_info}, we found that at least 50\% of the users identified as influential users by the SNIR model continue to be recognized as such by its dynamic extension. Regarding the impact of those identified as influential users, we find that their removal has the largest effect on information dissemination  in the dynamic SNIR model compared to the other four competitors, as seen the responses lost in Table \ref{dynamic:network_effect}.

\begin{CJK*}{UTF8}{gbsn}
\begin{table}[!h]
\caption{ Comparison of the identified influential users between the SNIR model and the dynamic SNIR model. Here, R, C, and L represent the sets identified by the SNIR model for reposts, comments, and likes, respectively. Likewise, R$^*$,  C$^*$, L$^*$ denote the sets identified by the dynamic SNIR model.  }
\centering
\begin{scriptsize}
\begin{tabular}{cccc}
\hline
User Name     & User Name In English& Identity  & Membership             \\ \hline
  &$Y_i$: log number of reposts ($|\mbox{R}|=10$, $|\mbox{R}^*|=11$) &&\\
  TFBOYS-易烊千玺 & Qianxi Yiyang        & super star &  R, R$^*$\\
  X玖少年团肖战DAYTOY & Zhan Xiao        & super star & R, R$^*$             \\
  BIGBIG张大大  & Dada Zhang & host           & R, R$^*$\\
  何炅          & Jiong He  & host           &  R, R$^*$\\
  韩国me2day      & me2day               & entertainment media   & R, R$^*$               \\
 于正          & Zheng Yu    & director and CEO & R, R$^*$\\
湖南卫视          & Hunan TV             & official media    & R, R$^*$               \\
THE9-虞书欣      & Shuxin Yu       & super star & R$^*$   \\
王嘉尔           & Team Wang   & super star & R$^*$ \\
河森堡           & Senbao He  & writer & R$^*$ \\
思想聚焦     & Focused Thinking & media & R$^*$ \\
 \hline
 &$Y_i$: log number of comments  ($|\mbox{C}|=9$, $|\mbox{C}^*|=10$)&&\\
 THE9-虞书欣      & Shuxin Yu       & super star & C, C$^*$      \\
  BIGBIG张大大     & Dada Zhang     & host       & C, C$^*$  \\
  何炅             & Jiong He       &host        & C, C$^*$   \\
  彭于晏           & Yuyan Peng     & actor      & C, C$^*$ \\
  于正            & Zheng Yu       & director and CEO & C, C$^*$ \\
     GAI周延         & Yan Zhou     & singer     & C, C$^*$  \\
     TFBOYS-易烊千玺 & Qianxi Yiyang        & super star &  C$^*$\\
     王嘉尔          & Team Wang   & super star & C$^*$ \\
     邓超           & Chao Deng   & actor & C$^*$\\
      张雪迎        & Xueying Zhang &  actress & C$^*$\\
    \hline
 &$Y_i$: log number of likes ($|\mbox{L}|=10$, $|\mbox{L}^*|=12$) &&\\
 TFBOYS-易烊千玺   & Qianxi Yiyang    & super star   &  L, L$^*$  \\
 BIGBIG张大大     & Dada Zhang     & host      & L, L$^*$  \\
 何炅             & Jiong He       &host        & L, L$^*$    \\
  GAI周延        & Yan Zhou       & singer     &   L, L$^*$   \\
  于正            & Zheng Yu       & director and CEO & L, L$^*$ \\
  王嘉尔           & Team Wang   & super star & L$^*$  \\
  THE9-虞书欣      & Shuxin Yu       & super star & L$^*$     \\
  思想聚焦     & Focused Thinking & media & L$^*$ \\
   X玖少年团肖战DAYTOY & Zhan Xiao        & super star & L$^*$             \\
    邓超           & Chao Deng   & actor & L$^*$\\
    河森堡           & Senbao He  & writer & L$^*$ \\
    郭德纲           & Degang Guo  & actor and CEO &L$^*$\\
     \hline
\end{tabular}
\end{scriptsize}
\label{dynamic:user_set_info}
\end{table}
\end{CJK*}

\begin{table}[!h]
\renewcommand\arraystretch{1.5}
\bc\emph{}
\caption{\label{dynamic:network_effect} Comparison among different methods: $\Delta$(R) represents the percentage of the responses lost due to the removal of influential users, while $\Delta$(F) represents the percentage of followers lost. }
\vspace{0.25cm}
{\scriptsize
\begin{tabular}{ccccccccccccccc}
\hline
 Response &  \multicolumn{2}{c}{Dynamic SNIR model}&&  \multicolumn{2}{c}{In degree-based} &&  \multicolumn{2}{c}{Response-based}&&  \multicolumn{2}{c}{Betweenness-based}&&  \multicolumn{2}{c}{Harmonic-based}\\
& $\Delta$(R) & $\Delta$(F) &&$\Delta$(R) & $\Delta$(F) &&$\Delta$(R) & $\Delta$(F) &&$\Delta$(R) & $\Delta$(F) &&$\Delta$(R) & $\Delta$(F) \\
 \cline{2-3}\cline{5-6}\cline{8-9}
 \cline{11-12}\cline{14-15}
  \\
  \hline
log(Reposts)&          74.9\%&      6.0\%&&     67.2\%&      7.5\%&&     69.7\%&      3.3\%&&     70.0\%&      6.6\%&&         63.5\%&      5.7\%\\
log(Comments)&     88.5\%&      5.5\%&&     82.9\%&      7.0\%&&     71.5\%&      3.0\%&&     84.1\%&      6.0\%&&         81.4\%&      5.1\%\\
log(Likes)&               90.1\%&      6.5\%&&     83.1\%&      8.0\%&&     74.6\%&      3.8\%&&     84.3\%&      6.9\%&&         80.0\%&      6.1\%\\
\hline
\end{tabular}}
\ec
\end{table}

\section{The choice  of candidate set $\mM$ in real data analysis}
\label{sec:choice}
 In the real data analysis, based on the theoretical requirement that $|\mM|\geq \lfloor N^{5/9}\rfloor$ and the in-degree histogram,  we directly used $|\mM|=\lfloor N^{2/3}\rfloor = 172$. To further study  the impact of the initial set $\mM$ on the selection results, we considered  two  additional choices for $|\mM|$, either  $125$ or $150$,  as suggested by the  in-degree histogram. The resulting influential user sets are summarized in  Table \ref{tab:user_info2}, in addition to those for when $|\mM|=172$.  When the response is log reposts, we find that the sets are identical when $|\mM|=\lfloor N^{2/3}\rfloor$ and $|\mM|=125$. These sets are identical with that selected by setting $|\mM|=150$, except for an additional user representing a renowned writer. For comments and likes, the estimated sets are the same whether $|\mM|=150$ or  $|\mM|=\lfloor N^{2/3}\rfloor$.  Interestingly, two extra users -- a superstar and the CEO of an entertainment company -- have been added to this set when $|\mM|=125$, possibly indicating that a smaller initial set may lead to the selection of more influential users to compensate for its limited size. We conclude that the outcomes are not very sensitive to the choice of $\mM$, as long as it is set reasonably large.


\begin{CJK*}{UTF8}{gbsn}
\begin{table}[!h]
\caption{Detailed information on the identified influential users for $|\mM|= \lfloor N^{2/3}\rfloor, 125$, and $150$, respectively. When the response is log reposts, we use R$_1$, R$_2$, and R$_3$ to denote the estimated sets for  $|\mM|=\lfloor N^{2/3}\rfloor$, $125$, and $150$, respectively. We define C$_j$ and L$_j$, $j=1,2,3$, likewise.}
\centering
\begin{scriptsize}
\begin{tabular}{cccc}
\hline
User Name     & User Name In English& Identity  & Membership             \\ \hline
  &$Y_i$: log number of reposts &&\\
  TFBOYS-易烊千玺 & Qianxi Yiyang        & super star &  R$_1$, R$_2$, R$_3$\\
   马伯庸           & Boyong Ma            & famous writer   & R$_1$, R$_2$, R$_3$              \\
  李荣浩        & Ronghao Li  & Singer       &  R$_1$, R$_2$, R$_3$\\
  X玖少年团肖战DAYTOY & Zhan Xiao        & super star & R$_1$, R$_2$, R$_3$             \\
  BIGBIG张大大  & Dada Zhang & host           & R$_1$, R$_2$, R$_3$\\
  何炅          & Jiong He  & host           &  R$_1$, R$_2$, R$_3$\\
  韩国me2day      & me2day               & entertainment media   & R$_1$, R$_2$, R$_3$               \\
 于正          & Zheng Yu    & director and CEO & R$_1$, R$_2$, R$_3$  \\
杨迪          & Di Yang   & host           &  R$_1$, R$_2$, R$_3$\\
湖南卫视          & Hunan TV             & official media    & R$_1$, R$_2$, R$_3$                \\
河森堡  & Senbao He   & Writer &  R$_3$\\

 \hline
 &$Y_i$: log number of comments &&\\
 THE9-虞书欣      & Shuxin Yu       & super star & C$_1$, C$_2$, C$_3$   \\
  李荣浩           & Ronghao  Li    & singer     & C$_1$, C$_2$, C$_3$  \\
  BIGBIG张大大     & Dada Zhang     & host       & C$_1$, C$_2$, C$_3$  \\
  何炅             & Jiong He       &host        & C$_1$, C$_2$, C$_3$  \\
  彭于晏           & Yuyan Peng     & actor      & C$_1$, C$_2$, C$_3$\\
  于正            & Zheng Yu       & director and CEO & C$_1$, C$_2$, C$_3$\\
   杨迪            & Di Yang       &host & C$_1$, C$_2$, C$_3$ \\
    GAI周延         & Yan Zhou     & singer     & C$_1$, C$_3$ \\
   Alex是大叔       & Alex         & writer     &  C$_1$, C$_3$ \\
   王嘉尔           & Team Wang    & super star &  C$_2$\\
   郭德纲           & Degang Guo  & actor and CEO &C$_2$\\
   \hline
 &$Y_i$: log number of likes &&\\
 TFBOYS-易烊千玺   & Qianxi Yiyang    & super star   & L$_1$, L$_2$, L$_3$\\
 李荣浩           & Ronghao  Li    & singer     & L$_1$, L$_2$, L$_3$  \\
 BIGBIG张大大     & Dada Zhang     & host      & L$_1$, L$_2$, L$_3$\\
 何炅             & Jiong He       &host        & L$_1$, L$_2$, L$_3$  \\
 彭于晏           & Yuyan Peng     & actor     & L$_1$, L$_2$, L$_3$   \\
  来去之间        & Gaofei Wang    & CEO  of Sina Weibo       & L$_1$, L$_2$, L$_3$\\
  于正            & Zheng Yu       & director and CEO & L$_1$, L$_2$, L$_3$\\
  杨迪           & Di Yang        &host & L$_1$, L$_2$, L$_3$\\
  GAI周延        & Yan Zhou       & singer     &  L$_1$, L$_3$ \\
  Alex是大叔       & Alex           & writer    & L$_1$, L$_3$ \\
  王嘉尔           & Team Wang    & super star &  L$_2$\\
  郭德纲           & Degang Guo  & actor and CEO &L$_2$\\
     \hline
\end{tabular}
\end{scriptsize}
\label{tab:user_info2}
\end{table}
\end{CJK*}

 Inspired by these results, in real practice, we suggest using several  values  of  $|\mM|$ after examining the in-degree histogram. We can follow either a majority rule by selecting users that appear in most sets, or an aggregate rule by taking their union. For the results presented in Tables \ref{tab:set_info}--\ref{tab:network_effect} in the main paper, we have opted for the former. We take this opportunity to evaluate the latter here to assess how the second rule might change any conclusion. Towards this, we present the identified important users in Table \ref{tab:user_info4} and the impact of removing these users in Table \ref{tab:approach22}. We can observe qualitatively similar patterns as those seen in Table \ref{tab:network_effect}.
\begin{CJK*}{UTF8}{gbsn}
\begin{table}[!h]
\caption{Detailed information for the identified influential users for approach 2. Followers: the number of followers, In-degree: the in-degree value, Likes: the number of likes, Comments: the number of Comments, Reposts: the number of reposts, Membership: R for reposts, C for comments and L for likes. }
\centering
\begin{scriptsize}
\begin{tabular}{c|c|c|c|c|c|c|c}
\hline
User Name     & Name in English & Followers    & In-degree  & Likes   & Comments & Reposts & Membership             \\ \hline
BIGBIG张大大     & Dada Zhang           & 32441000  & 179  & 145086  & 9119     & 93294   & R, C, L \\ \hline
何炅            & Jiong He             & 120000000 & 667  & 142891  & 5327     & 15418   & R, C, L\\ \hline
李荣浩           & Ronghao Li           & 23537109  & 205  & 10613   & 760      & 1065    & R, C, L \\ \hline
杨迪            & Di Yang              & 7973000   & 223   & 27324   & 2560     & 317     & R, C, L \\ \hline
于正            & Zheng Yu             & 8176000   & 236    & 872     & 72       & 42      & R, C, L \\ \hline
TFBOYS-易烊千玺   & Qianxi Yiyang        & 88623341  & 349  & 1397934 & 125832   & 1109797 & R, L        \\ \hline
彭于晏           & Yuyan Peng           & 32944000  & 242   & 13601   & 1256     & 1245    & C, L         \\ \hline
Alex是大叔       & Alex                 & 17873113  & 104   & 483     & 40       & 387     & C, L       \\ \hline
GAI周延         & Yan Zhou             & 11379901  & 106    & 67597   & 16550    & 26437   & C,  L        \\ \hline
 王嘉尔           & Team Wang          &  29472756 &  240   & 265883  & 158216   & 258592  & C,  L           \\ \hline
 郭德纲           & Degang Guo         & 73352000  & 189    & 49066     &5         & 1240  & C,  L           \\ \hline
X玖少年团肖战DAYTOY & Zhan Xiao        & 29547091  & 201  & 5389162 & 1000000  & 1000000 & R               \\ \hline
韩国me2day      & me2day               & 26204000  & 145   & 1155    & 56       & 310     & R               \\ \hline
湖南卫视          & Hunan TV             & 15996639 & 210  & 823     & 437      & 156     & R               \\ \hline
马伯庸           & Boyong Ma            & 7779218   & 184 & 607     & 51       & 1262    & R              \\ \hline
河森堡           & Senbao He            & 6369000  &  140 & 214      &19        &225     &R             \\ \hline
THE9-虞书欣      & Shuxin Yu            & 15724775  & 138 & 78847   & 26752    & 13861   & C               \\ \hline
来去之间         & Gaofei Wang          & 768000   & 243  & 339     & 73       & 1564    & L              \\ \hline
\end{tabular}
\end{scriptsize}
\label{tab:user_info4}
\end{table}
\end{CJK*}

\begin{table}[!h]
\renewcommand\arraystretch{1.5}
\bc\emph{}
\caption{\label{tab:approach22} The  comparison among different methods  for  approach 2. $\Delta$(R) is the percentage of the responses lost due to the removal of influential users, while $\Delta$(F) is the percentage of followers lost.}
\vspace{0.25cm}
{\scriptsize
\begin{tabular}{ccccccccccccccc}
\hline
 Response &  \multicolumn{2}{c}{SNIR under approach 2}&&  \multicolumn{2}{c}{In degree-based} &&  \multicolumn{2}{c}{Response-based}&&  \multicolumn{2}{c}{Betweenness-based}&&  \multicolumn{2}{c}{Harmonic-based}\\
& $\Delta$(R) & $\Delta$(F) &&$\Delta$(R) & $\Delta$(F) &&$\Delta$(R) & $\Delta$(F) &&$\Delta$(R) & $\Delta$(F) &&$\Delta$(R) & $\Delta$(F) \\
 \cline{2-3}\cline{5-6}\cline{8-9}
 \cline{11-12}\cline{14-15}
  \\
  \hline
log(Reposts)     &     76.0\%&      6.0\%&&     67.2\%&      7.5\%&&     69.7\%&      3.3\%&&     70.0\%&      6.6\%&&       63.5\%&      5.7\%\\
log(Comments)&     88.6\%&      5.6\%&&     83.1\%&      7.5\%&&     68.1\%&      3.0\%&&     85.4\%&      6.6\%&&        81.6\%&      5.7\%\\
log(Likes)          &     90.1\%&      6.5\%&&     83.1\%&      8.0\%&&     74.6\%&      3.8\%&&     84.3\%&      6.9\%&&        80.0\%&      6.1\%\\
\hline
\end{tabular}}
\ec
\end{table}

\section{An extension of the model to include covariates and heterogenous random error}\label{sec:withX}
This section serves two purposes. First, we assess the robustness of  our approach  when the errors in
SNIR are heteroskedastic. Second, we discuss an extension of  the SNIR model to include covariates.
Formally, this extended model is specified as
\beqr
Y_i =\mu_i+\sum_{j = 1}^N \rho_j a_{ij} Y_j +Z_i^\top \beta+ \varepsilon_i,  \quad i=1,\cdots, N, \n
\label{SNIRX}
\eeqr
where  $Z_i\in\mR^p$ is the exogenous covariate of user $i$ satisfying $E(Z_i|\varepsilon_i)=0$
and
 $\varepsilon_i$'s are independent  with mean 0 and heterogeneous variance $\sigma_i^2$.
 We consider the three different network structures specified  in Section 4.1 and generate
  $\varepsilon_i$  independently  from  $N(0,\sigma_i^2)$ with
 $\sigma_i$ from  a uniform distribution $U(0.5,1.5)$. We independently generate each $Z_i$ from $N(0,\Sigma)$ with $\Sigma=(\sigma_{ij})$ and $\sigma_{ij}=0.5^{|i-j|}$.
 Since  $Z_i$ is  exogenous,  we can profile the effect of $Z$ by defining
   $\mY^*= \mY- \mZ \hat \beta$, where $\mZ=(Z_1,\cdots, Z_N)^\top\in \mR^{N\times p}$ and $ \hat \beta= (\mZ^\top \mZ)^{-1} \mZ^\top\mY$.
 Afterwards, we apply the forward-addition algorithm  to $(\mY^*, A)$ to obtain the associated estimator for
 $\mS_1$ and $\bm \rho$.
 The detailed results are summarized in Table \ref{tab:tS1} below, which are qualitatively similar to the results in Tables \ref{tab:t0}--\ref{tab:real_network}. The results imply further that our estimation method is robust  with respect to heterogeneous  random errors.

\section{Proof of Theorem \ref{thm:E.1} and Theorem \ref{thm:E.2}}
\label{sec:thmE.1}
In this section, we add two new theorems.
Theorem \ref{thm:E.1} states that  the difference between $Q(\bm \rho_{(\mS_{1})})$ and $\wt  Q(\bm \rho_{(\mS_{1})})$ is ignorable asymptotically, making the CMLE $\bm{\hat \rho}_{(\mS_1)}$ as efficient as FMLE.
Theorem \ref{thm:E.2} is about  the  asymptotic  normality of $\hat {\bm\rho}_{(\mS_1)}$.
It collaborates with Corollary \ref{corollary 1} when $\mS^{(k^*)}=\mS_1$.

\bet
\label{thm:E.1}
 Under  Conditions (C1)–(C4) discussed in Appendix \ref{sec:condition}, we have
$\{\wt  Q(\mu, \bm \rho_{(\mS_{1})})-Q(\bm \rho_{(\mS_{1})})\}/Q(\bm \rho_{(\mS_{1})})=O_p(N^{\alpha_1-1}).$
\eet
{\bf Proof.} By Condition (C2), one can verify that the eigenvalues of $H_1^\top H_1$ are all bounded from 0 and infinity.
Therefore, we have $N^{-1} Q(\bm \rho_{(\mS_{1})})\rightarrow_p \sigma^2>0$. On the other hand,  we have
\[(\mY_{(\mS_1)}- H_1^{-1}  \mu \bm 1_{(\mS_1)})^\top H_1^\top H_1 (\mY_{(\mS_1)}- H_1^{-1}  \mu \bm 1_{(\mS_1)})\]
\[\leq \lambda_{\max}(H_1^\top  H_1) (\mY_{(\mS_1)}- H_1^{-1}  \mu \bm 1_{(\mS_1)})^\top (\mY_{(\mS_1)}- H_1^{-1}  \mu \bm 1_{(\mS_1)}).\]
By Condition (C2), we know that $\lambda_{\max}(H_1^\top  H_1)$  is bounded and the above inequality can   be further bounded by $O_p(|\mS_1|)$.
Together with the fact that  $\ln |H_1|=O_p(|\mS_1|)$, we have
$\wt Q(\mu, \bm \rho_{(\mS_{1})})=O_p(N)$
and  $ \wt Q(\mu, \bm \rho_{(\mS_{1})})-Q( \bm \rho_{(\mS_{1})})=O_p(N^{\alpha_1})$.
Consequently, the result of Theorem \ref{thm:E.1} is obviously satisfied.
\bet
\label{thm:E.2}
(Asymptotic normality of $\hat {\bm\rho}_{(\mS_1)}$)
Under Conditions (C1)--(C4)  and (C7), we have
$\hat {\bm\rho}_{(\mS_{1})} $  is asymptotically normal, in the sense that
 \[|\mS_{1}|^{-1/2}N^{1/2}\Psi \Lambda_1^{1/2}\mbox{diag}(\mY_{(\mS_1)})(\hat {\bm\rho}_{(\mS_{1})}-\bm \rho_{(\mS_{1})}) \xrightarrow{d} N(0,G) \]
 as $N \rightarrow \infty$, where $\Psi$ is an arbitrary  $q \times |\mS_{1}|$  matrix with $q<\infty$ such that $|\mS_{1}|^{-1}\Psi\Psi^\top \rightarrow G $ for some
 $q \times q $ nonnegative symmetric matrix $G$, and $N(0,G)$ is the multivariate normal distribution with mean $0$ and covariance matrix $G$.
\eet
{\bf Proof.}  Recall that $\Lambda_1\in\mR^{|\mS_{1}|\times |\mS_{1}|}$  is a finite positive definite matrix defined in Condition (C4)  with
 $\|N^{-1}A_{(\mS_{0},\mS_{1})}^{\top} A_{(\mS_{0},\mS_{1})}-\Lambda_1\|_F=o(1)$.
Without loss of generality, consider $q=1$. The proof for general $q>1$ is similar.
Recall that $\bar\mX_{(\mS_{1})} = A_{(\mS_0,\mS_{1})} \mbox{diag} ( \mY_{(\mS_{1})} )$. We
have
\[N^{1/2}\mbox{diag}(\mY_{(\mS_1)})\big(
\bm{\hat \rho}_{(\mS_{1})}-\bm {\rho}_{(\mS_{1})}\big)=
N^{1/2}\mbox{diag}(\mY_{(\mS_1)})\{\bar\mX_{(\mS_{1})}^{\top}\bar\mX_{(\mS_{1})}\}^{-1} \bar\mX_{(\mS_{1})}^{\top}\mE_{(\mS_0)}\]
\[=
N^{1/2}\mbox{diag}(\mY_{(\mS_1)}) \Big\{\mbox{diag}(\mY_{(\mS_1)}) A_{(\mS_0,\mS_{1})}^\top A_{(\mS_0,\mS_{1})} \mbox{diag}(\mY_{(\mS_1)})\Big\}^{-1} \mbox{diag}(\mY_{(\mS_1)}) A_{(\mS_0,\mS_{1})}^\top \mE_{(\mS_0)}
\]
\[
=N^{-1/2} \big\{N^{-1} A_{(\mS_0,\mS_{1})}^\top A_{(\mS_0,\mS_{1})}\big\}^{-1} A_{(\mS_0,\mS_{1})}^\top \mE_{(\mS_0)}.\]

Subsequently, by Condition (C4),  we have
\[|\mS_1|^{-1/2} N^{1/2}\Psi \Lambda_1^{1/2}
\mbox{diag}(\mY_{\mS_1})
\Big(\hat{\bm\rho}_{(\mS_1)}
-{\bm \rho}_{(\mS_{1})}\Big)=
|\mS_1|^{-1/2}\Psi N^{-1/2} \Lambda_1^{-1/2} A_{(\mS_0,\mS_{1})}^\top \mE_{(\mS_0)}\big\{1+o_p(1)\big\}\]
\[=|\mS_1|^{-1/2}\Psi\Omega\mE_{(\mS_0)}\big\{1+o_p(1)\big\}=\mU\mE_{(\mS_0)}\{1+o_p(1)\},\]
where  $\Omega=N^{-1/2} \Lambda_1^{-1/2} A_{(\mS_0,\mS_{1})}^\top \in\mR^{|\mS_1|\times |\mS_0|}$,
 $\mU=|\mS_1|^{-1/2} \Psi\Omega$ and the term $o_p(1)$ is uniformly across all dimension of $\mE_{(\mS_0)}$.
Let  $\mU=(u_1,\cdots, u_{|\mS_0|})\in\mR^{1\times|\mS_0|}$ and note that $\mE_{(\mS_0)}=(\varepsilon_j: j\in\mS_0)^\top
\in\mR^{|\mS_0|}$.
We have $\mU \mE_{(\mS_0)}= \sum_{j\in\mS_0} u_j\varepsilon_j$. Define
$S^2_N= \sum_{j\in\mS_0}  \var(u_j\varepsilon_j)= \sum_{j\in\mS_0}  u_j^2$.
By the  Lyapunov central limit  theorem, it suffices to show
\beqr
\lim_{N\rightarrow \infty}  \frac{1}{S^4_N}  \sum_{j\in \mS_0} E(u_j\varepsilon_j)^4=0.\label{A.1}
\eeqr
By Condition (C1), we know that $E(\varepsilon_j^4)<\infty$. To prove \eqref{A.1}, it suffices to prove that
\[\lim_{N\rightarrow \infty}  \sum_{j\in\mS_0} u_j^4/ \big(\sum_{j\in\mS_0} u_j^2\big)^{2} =0.\]
We first evaluate
$(\sum_{j\in\mS_0} u_j^2)^2 $ and
we can verify that
\[(\sum_{j\in\mS_0} u_j^2)^2= \parallel \mU\parallel^4=(|\mS_1|^{-1} \Psi \Psi^\top)^2\big\{1+o(1)\big\} =O(1),\]
and that it is bounded away  from 0.

Next consider the term $\sum_{j\in\mS_0} u_j^4$.  Define $\Omega=(\Omega_{ij})$. By the Cauchy-Schwartz inequality, we have
\[\sum_{j\in\mS_0} u_j^4= |\mS_1|^{-2} \sum_{j\in\mS_0} (\sum_{i\in\mS_1} \Psi_i \Omega_{ij})^4\leq  |\mS_1|^{-2} \sum_{j\in\mS_0}
\Big\{\big(\sum_{i\in\mS_1}\Psi^2_i\big) \big(\sum_{i\in\mS_1}\Omega_{ij}^2\big)\Big\}^2 \]
\[\leq  C_{\Psi}\sum_{j\in\mS_0} \Big(\sum_{i\in\mS_1}\Omega_{ij}^2\Big)^2\leq
C_{\Psi} |\mS_1|^2 \max_i \sum_{j\in\mS_0}\Omega_{ij}^4=O(|\mS_1|^2 N^{-\alpha_3})=o(1),\]
where  $C_{\Psi}$ is   a finite positive constant.  The last two equalities of the above equation is implied by Condition (C7).   Combining the above results, we have proven \eqref{A.1}, which completes the proof of the theorem. 

\section{Proof of Theorem \ref{theorem3.1}(1)}\label{sec:thm3.1}

{\bf Proof.} By definition, we have $\mS_1\subset \mM$. By  working model \eqref{work_model},
$\mY_{(\mM^c)} =  A_{(\mM^c,\mM)} \mD_{(\mM)}  \mY_{(\mM)}  + \mE_{(\mM^c)}$, where $\mE_{(\mM^c)}$ is the corresponding subset of $\mE_{(\mS_0)}$.
To prove the theorem result,
similar to Wang (2009), for any given subset $\mS^{(k)}\not\in\mS_1$ and $\mS^{(k+1)}=\mS^{(k)}\cup a_k$, it suffices to
prove
\[\Theta(k)=\RSS(\mS^{(k)})-\RSS(\mS^{(k+1)})\geq C_K N^{1-(m-\alpha_1)}\]
for a finite positive constant $C_K$ if $a_k\not\in\mS_1$ uniformly for any $k\leq O(N^m)$.
Accordingly, a relevant user will be selected within at least every $O(N^{m-\alpha_1})$ steps. Since $|\mS_1|=O(N^{\alpha_1})$ by
Condition (C3), then all the relevant users will be selected within $O(N^{m})$ steps.

By definition, we have
$\Theta(k)=\|H_{a_k}^{(k)}\mQ_{\mS^{(k)}}\mY_{(\mM^c)}\|^2$ with $\mQ_{\mS^{(k)}}=I-\mX_{(\mS^{(k)})}(\mX_{(\mS^{(k)})}^\top \mX_{\mS^{(k)})})^{-1} \mX_{(\mS^{(k)})}^\top$,
$H_j^{(k)}=\mX_{(j)}^{(k)}\mX_{(j)}^{(k)\top}\|\mX_{(j)}^{(k)}\|^{-2}$, $\mX_{(j)}^{(k)}=\mQ_{\mS^{(k)}}\mX_{(j)}$ and
$\mX_{(j)}=A_{(\mM^c, j)} Y_j$.
 Since $a_k\not\in\mS_1$ and $a_k$ has been selected in the $(k+1)$-th step, we have
\[\Theta(k)\geq \max_{j\in\mS_1}\Big\|H_j^{(k)}\mQ_{\mS^{(k)}}\mY_{(\mM^c)} \Big\|^2=\max_{j\in\mS_1} \Big\|H_j^{(k)}\mQ_{\mS^{(k)}}\big(\mX_{(\mS_{1})}\bm \rho_{(\mS_{1})}+\mE_{(\mM^c)}\big)\Big\|^2\]
\[\geq \max_{j\in\mS_1}\Big\|H_j^{(k)}\mQ_{\mS^{(k)}}\mX_{(\mS_{1})}\bm\rho_{(\mS_{1})}\Big\|^2-\max_{j\in\mS_{1}}\Big\|H_j^{(k)}\mQ_{\mS^{(k)}}\mE_{(\mM^c)}\Big\|^2.\]
We next consider the above two parts separately in the following two steps.

{\sc Step I.} We first consider  $\max_{j\in\mS_{1}}\|H_j^{(k)}\mQ_{\mS^{(k)}}\mE_{(\mM^c)}\|^2$.
By  the fact that
$H_j^{(k)}$ is an
idempotent matrix, we have
$$\max_{j\in\mS_{1}}\|H_j^{(k)}\mQ_{\mS^{(k)}}\mE_{(\mM^c)}\|^2=\max_{j\in\mS_{1}}\|\mX_{(j)}^\top\mQ_{\mS^{(k)}}\mE_{(\mM^c)}\|^2/\|\mX_{(j)}^{(k)}\|^2.$$
We evaluate the order of $\|\mX_{(j)}^{(k)}\|^2$. Note that
$$\mX_{(j)}^{(k)}=\mQ_{\mS^{(k)}}\mX_{(j)}=\mX_{(j)}-
\mX_{(\mS^{(k)})}\big\{\mX_{(\mS^{(k)})}^\top\mX_{(\mS^{(k)})}\big\}^{-1}\mX_{(\mS^{(k)})}^\top\mX_{(j)}=Y_j\big\{A_{(\mM^c,j)}-A_{(\mM^c,\mS^{(k)})}\theta_{j,\mS^{(k)}}\big\},$$
where $\theta_{j,\mS^{(k)}}=\big\{A_{(\mM^c,\mS^{(k)})}^\top A_{(\mM^c, \mS^{(k)})}\big\}^{-1}A_{(\mM^c, \mS^{(k)})}^\top A_{(\mM^c, j)}$.
Employing Condition (C6)  on the set $\mS^{(k)}\cup \mS_1$,  we have that
\[\Big\|\mX_{(j)}^{(k)} \Big\|^2=Y_j^2 (1,-\theta_{j,\mS^{(k)}}^\top) A_{(\mM^c,j\cup\mS^{(k)})}^\top A_{(\mM^c,j\cup\mS^{(k)})}(1,-\theta_{j,\mS^{(k)}}^\top)^\top\]
\[\geq \tau_{\min,2} \min_{i\in\mM} |\mN_i| Y_j^2\geq \tau_{\min,2} N^{\alpha_2} Y_j^2/C_3\]
with probability approaching 1 uniformly for any $j\in\mS_1$.
Consequently, we have
\[\max_{j\in\mS_{1}}\Big\|H_j^{(k)}\mQ_{\mS^{(k)}}\mE_{(\mM^c)} \Big\|^2\leq C_3\tau_{\min,2}^{-1}N^{-\alpha_3}\max_{j\in\mS_{1}}\Big\|\mX_{(j)}^\top\mQ_{\mS^{(k)}}\mE_{(\mM^c)} \Big\|^2/Y_j^2.\]
In addition, note that
$$\Big\|\mX_{(j)}^\top\mQ_{\mS^{(k)}}\mE_{(\mM^c)}\Big\|^2=\mE_{(\mM^c)}^\top\mQ_{\mS^{(k)}} \mX_{(j)}\mX_{(j)}^\top \mQ_{\mS^{(k)}} \mE_{(\mM^c)},$$
and \[\mbox{tr}\Big\{\mQ_{\mS^{(k)}} \mX_{(j)}\mX_{(j)}^\top \mQ_{\mS^{(k)}}\Big\}=
\mbox{tr}\Big\{\mX_{(j)}^\top \mQ_{\mS^{(k)}}\mX_{(j)}\Big\}
\leq \mbox{tr}\{\mX_{(j)}\mX_{(j)}^\top\}=Y_j^2 A_{(_{(\mM^c)}, j)}^\top A_{(_{(\mM^c)}, j)}.\]
Accordingly, $\|\mX_{(j)}^\top\mQ_{\mS^{(k)}}\mE_{(\mM^c)}\|^2/Y_j^2$
is a quadratic form of $\mE_{(\mM^c)}$ with mean smaller than
$A_{(_{(\mM^c)}, j)}^\top A_{(_{(\mM^c)}, j)}\leq \tau_{\max,1} N$.
Consequently, by the Hanson-Wright inequality on quadratic forms (Hanson and Wright, 1971; Wright, 1973), together with
the Bonferroni inequality, we have
\[\max_{j\in\mS_1} \max_{|\mS^{(k)}|\leq m^*}\Big\|\mX_{(j)}^\top\mQ_{\mS^{(k)}}\mE_{(\mM^c)}\Big\|^2/Y_j^2 =O(m^* N \log N),\]
 which immediately leads to
\beq
\max_{j\in\mS_{1}}\Big\|H_j^{(k)}\mQ_{\mS^{(k)}}\mE_{(\mM^c)}\Big\|^2\leq C_{K_1}N^{1-\alpha_2+m}\log N
\label{step1}
\eeq
for a finite positive constant $C_{K_1}$.

{\sc Step II.} We next consider $\max_{j\in\mS_{1}}\|H_j^{(k)}\mQ_{\mS^{(k)}}\mX_{(\mS_{1})}\bm\rho_{(\mS_{1})}\|^2$.
By the proof of Step I, we have
$\|\mX_{(j)}^{(k)}\|^2\leq \|\mX_{(j)}\|^2\leq \tau_{\max,1} NY_j^2$, which is implied  by Condition (C4).
Subsequently,
\[\max_{j\in\mS_{1}}\Big\|H_j^{(k)}\mQ_{\mS^{(k)}}\mX_{(\mS_{1})}\rho_{(\mS_{1})} \Big\|^2=
\max_{j\in\mS_{1}}\Big\|\mX_{(j)}^{(k)}\Big\|^{-2}\Big\|\mX_{(j)}^\top \mQ_{\mS^{(k)}}\mX_{(\mS_{1})}\bm\rho_{(\mS_{1})}\Big\|^2\]
\beq \geq \tau_{\max,1}^{-1} N^{-1} \max_{j\in\mS_{1}}\Big\|\mX_{(j)}^\top\mQ_{\mS^{(k)}}\mX_{(\mS_{1})}\bm\rho_{(\mS_{1})}\Big\|^2/Y_j^2.\label{6.3}\eeq
In addition, by the Cauchy-Schwarz inequality, we have
\[\Big\|\mQ_{\mS^{(k)}}\mX_{(\mS_{1})}\bm\rho_{(\mS_{1})}\Big\|^2=\bm\rho_{(\mS_{1})}^\top \mX_{(\mS_{1})}^\top \mQ_{\mS^{(k)}} \big\{\mX_{(\mS_{1})}\bm\rho_{(\mS_{1})}\big\}
=\sum_{j\in\mS_1} \rho_j\mX_{(j)}^\top\mQ_{\mS^{(k)}} \big\{\mX_{(\mS_{1})}\bm\rho_{(\mS_{1})}\big\}\]
\[\leq \big\|\bm\rho\big\| \Big[\sum_{j\in\mS_{1}} \big\{\mX_{(j)}^\top\mQ_{\mS^{(k)}}\mX_{(\mS_{1})}\bm\rho_{(\mS_{1})}\big\}^2\Big]^{1/2}
\leq |\mS_{1}|^{1/2} \big\|\bm\rho \big\|\max_{j\in\mS_{1}}\Big\|\mX_{(j)}\mQ_{\mS^{(k)}}\mX_{(\mS_{1})}\bm\rho_{(\mS_{1})}\Big\|.\]
By Condition (C5), we have $\|\bm\rho\|\leq  \rho_{\max} |\mS_{1}|^{1/2}$, then

\[\max_{j\in\mS_{1}}\Big\|\mX_{(j)}\mQ_{\mS^{(k)}}\mX_{(\mS_{1})}\bm\rho_{(\mS_{1})}\Big\|\geq \rho_{\max}^{-1}|\mS_{1}|^{-1}
\Big\|\mQ_{\mS^{(k)}}\mX_{(\mS_{1})}\bm\rho_{(\mS_{1})}\Big\|^2.\]
Plugging in this result into the above inequality \eqref{6.3}, we then have
\beq \max_{j\in\mS_{1}}\Big\|H_j^{(k)}\mQ_{\mS^{(k)}}\mX_{(\mS_{1})}\bm\rho_{(\mS_{1})}\Big\|^2\geq
\tau_{\max,1}^{-1}N^{-1}\bm\rho_{\max}^{-2}|\mS_{1}|^{-2}
\Big\|\mQ_{\mS^{(k)}}\mX_{(\mS_{1})}\bm\rho_{(\mS_{1})}\Big\|^4 \big/\max_{j\in\mS_{1}}Y_j^2.\label{6.4} \eeq
By Condition (C2), one can verify that $\max_{j\in\mS_{1}}Y_j^2=O_p(\log N)$
by the Bonferroni inequality and the Hanson-Wright inequality on quadratic forms \citep{hanson1971bound, wright1973bound}.
In addition, define $\zeta_{(\mS^{(k)})}=\{\mX_{(\mS^{(k)})}^\top \mX_{(\mS^{(k)})}\big\}^{-1}\mX_{(\mS^{(k)})}^\top\mX_{(\mS_{1})}\bm\rho_{(\mS_{1})}$.
Then,
by Conditions (C4) and (C5), we have
\beq
\Big\|\mQ_{\mS^{(k)}}\mX_{(\mS_{1})}\bm\rho_{(\mS_{1})} \Big\|^2=
\Big\|\mX_{(\mS_{1})}\bm\rho_{(\mS_{1})}-\mX_{(\mS^{(k)})}\zeta_{(\mS^{(k)})}\Big\|^2\geq \mN_{\min} \tau_{\min,2}\rho_{\min}^2\min_{j\in\mS_1} Y_j^2.\label{6.5}
\eeq
By Conditions (C1) and (C2), together with the Bonferroni inequality, we can obtain that $\min_{j\in\mS_1}Y_j^2=O_p\{|\mS_1|^{-1}\}=O_p(N^{-\alpha_1})$.
Plugging in this result into \eqref{6.3} and \eqref{6.4}, and by Condition (C5), we have
\beq
\max_{j\in\mS_{1}}\Big\|H_j^{(k)}\mQ_{\mS^{(k)}}\mX_{(\mS_{1})}\bm\rho_{(\mS_{1})}\Big\|^2\geq C_{K_2} N^{2\alpha_2-1-4\alpha_1}/\log N
\label{6.6}
\eeq
for a finite positive constant $C_{K_2}$.
Combining the results in \eqref{step1} and \eqref{6.6}, we have
\[\Theta(k)\geq C_{K_3} N^{2\alpha_2-1-4\alpha_1}/\log N-C_{K_2}N^{1-\alpha_3+m}\log N \geq 2^{-1}C_{K_3} N^{2\alpha_2-1-4\alpha_1-\eta} \]
for any arbitrary small constant $\eta>0$,
where the above inequality holds since $m=2+5\alpha_1-2\alpha_2$
and by Condition
(C6) that $4+9\alpha_1<5\alpha_2$. Then,
we have
 $\Theta(k)\geq C_K N^{1+\alpha_1-m}$ for a finite positive constant $C_K$.
 Then, with probability approaching 1, we have
 $$\Theta(k)\geq 2^{-1}C_KN^{1+\alpha_1-m}.$$
Analogue to the discussion in Wang (2009), a relevant predictor will be selected within at least every $O(N^{m-\alpha_1})$ steps. Since $|\mS_{1}|=O(N^{\alpha_1})$ by
Condition (C2), then all the relevant predictors will be selected within $O(N^{m})=O(N^{2+5\alpha_1-2\alpha_2})$ steps, which completes the proof of this theorem.

\section{Proof of Theorem \ref{theorem3.1}(2)}\label{sec:thm3.3}
{\bf Proof.}
Define $k_{\min}=\min\{k:\mS_1\in \mS^{(k)}\}$.
We know that $k_{\min}$ is well defined and satisfies $k_{\min}=O(N^{2+5\alpha_1-2\alpha_2})$ with probability approaching one by the result
of Theorem \ref{theorem3.1}(1). Hence, analogous to Wang (2009), to prove the theorem, it suffices to show
$$P\Big(\min_{k<k_{\min}} \big\{\mbox{EBIC}(\mS^{(k)})-\mbox{EBIC}(\mS^{(k+1)})\big\}>0\Big)\rightarrow 1.$$
By definition, we obtain
\beq
\begin{aligned}
\mbox{EBIC}(\mS^{(k)})-\mbox{EBIC}(\mS^{(k+1)})
&=\log\Big\{\frac{\RSS(\mS^{(k)})}{\RSS(\mS^{(k+1)})}\Big\}
-|\mM^c|^{-1} \Big\{\log(|\mM^c|)+2\log(|\mM|)\Big\}+o(1)
\n \\
&\geq \log\Big\{1+\frac{\RSS(\mS^{(k)})-\RSS(\mS^{(k+1)})}{\RSS(\mS^{(k+1)})}\Big\}
-3|\mM^c|^{-1} \log(|\mM|).
\end{aligned}
\eeq
By the results of Theorem \ref{theorem3.1}(1), we have
$\RSS(\mS^{(k)})-\RSS(\mS^{(k+1)})\geq C_K N^{1+\alpha_1-m}$ and  $\RSS(\mS^{(k+1)})\leq \|\mY\|^2\leq C_S N$ for two finite constants $C_K$ and $C_S$.
Accordingly, with probability tending to one, the right hand side of the above equation  is no less than
\[
\log\Big\{1+C_KN^{\alpha_1-m}/C_S\Big\}-3(|\mM^c|)^{-1} \log(|\mM|)\]
\[\geq \min\Big\{\log 2, 2^{-1}C_K N^{\alpha_1-m}/C_S\Big\} -3N^{-m}  \log(m^*),
\]
where the first inequality is implied by $\log(1+x)\geq \min\{\log 2, 0.5x\}$ for any $x>0$.
In addition, by Condition (C2) and (C6), we have $\alpha_1$ and $m$  are  positive constants, which immediately leads to
$2^{-1}C_K N^{\alpha_1-m}/C_S-3N^{-m}  \log(m^*)>0$.
Furthermore, we obtain $\log 2-3N^{-m}  \log(m^*)>0$ when $N$ is sufficiently large.
Combining the results above, we have completed the proof.

\section{Proof of Corollary \ref{corollary 1} }\label{sec:corollary}
{\bf Proof.}
Analogous to the proof of Theorem \ref{thm:E.2},
we use  $q=1$ for illustration. The
 proof for general $q>1$ is similar. Note that $\mS^{(k^*)}$ is the selected
model that minimizes EBIC. For notation simplicity, we use generic notation $\mS$ to denote $\mS^{(k^*)}$ in the reminder of this proof.
Accordingly, ${\bar \mX}_{(\mS^{(k^*)})} = A_{(\mS^c,\mS)} \mbox{diag} ( \mY_{(\mS)} )$.
We have
\[B^{1/2}_N(\mS)\mbox{diag}(\mY_{(\mS)})(
\bm{\hat \rho}_{(\mS)}-\bm \rho_{(\mS)})=
B^{1/2}_N(\mS)\mbox{diag}(\mY_{(\mS)})\big\{{\bar \mX}_{(\mS)}^{\top}{\bar \mX}_{(\mS)}\big\}^{-1} {\bar \mX}_{(\mS)}^{\top}\mE_{(\mS^c)}\]
\[=
B^{1/2}_N(\mS)\mbox{diag}(\mY_{(\mS)}) \Big\{\mbox{diag}(\mY_{(\mS)}) A_{(\mS^c,\mS)}^\top A_{(\mS^c,\mS)} \mbox{diag}(\mY_{(\mS)})\Big\}^{-1} \mbox{diag}(\mY_{(\mS)}) A_{(\mS^c,\mS)}^\top \mE_{(\mS^c)}\]
\[=B^{-1/2}_N(\mS)\big\{B^{-1}_N(\mS) A_{(\mS^c,\mS)}^\top A_{(\mS^c,\mS)}\big\}^{-1} A_{(\mS^c,\mS)}^\top \mE_{(\mS^c)}.\]
Subsequently,  $|\mS|^{-1/2} \wt\Psi B^{1/2}_N(\mS) \Lambda^{1/2}(\mS)\mbox{diag}(\mY_{\mS})
({\widehat {\bm\rho}}_{(\mS)}-\bm \rho_{(\mS)})$ can be expressed as
$$|\mS|^{-1/2}\wt\Psi \big\{ \Lambda(\mS)B_N(\mS)\big\}^{-1/2} A_{(\mM^c,\mS)}^\top \mE_{(\mM^c)}\{1+o_p(1)\}
=|\mS|^{-1/2}\wt\Psi \wt\Omega\mE_{(\mS^c)}\{1+o_p(1)\}=\wt\mU\mE_{(\mS^c)}\{1+o_p(1)\},$$
where  $\wt\Omega=\{ \Lambda(\mS)B_N(S)\}^{-1/2} A_{(\mS^c,\mS)}^\top$,
 $\wt\mU=|\mS|^{-1/2} \wt\Psi \wt\Omega$.
Let  $\wt\mU=(\wt u_1,\cdots, \wt u_{|\mS^c|})\in\mR^{1\times|\mS^c|}$ and note that $\mE_{(\mS^c)}=(\varepsilon_j: j\in \mS^c)^\top
\in\mR^{|\mS^c|}$.
We have $\wt\mU \mE_{(\mS^c)}= \sum_{j\in\mS^c} \wt u_j\varepsilon_j$. Define
$\wt S^2_N= \sum_{j\in\mS^c}  \var(\wt u_j\varepsilon_j)= \sum_{j\in\mS^c} \wt u_j^2$.
By the  Lyapunov central limit  theorem, it suffices to show
\beqr
\lim_{N\rightarrow \infty}  \frac{1}{\wt S^4_N}  \sum_{j\in \mS^c} E(\wt u_j\varepsilon_j)^4=0.\label{E.1}
\eeqr
 To prove \eqref{E.1}, it suffices to prove that
\[\lim_{N\rightarrow \infty}  \sum_{j\in\mS^c} \wt u_j^4/ \big(\sum_{j\in\mS^c} \wt u_j^2\big)^{2} =0.\]
One can verify that
\[(\sum_{j\in\mS^c} \wt u_j^2)^2= \parallel \wt \mU\parallel^4=(|\mS|^{-1} \wt \Psi \wt \Psi^\top)^2\big\{1+o(1)\big\} =O(1),\]
and it is bounded away from 0.
In addition, by Condition (C7), we have that there exists a finite positive constant $C_{\wt\Psi}$ such that
\[\sum_{j\in\mS^c} \wt u_j^4= |\mS|^{-2} \sum_{j\in\mS^c} (\sum_{i\in\mS} \wt \Psi_i \wt\Omega_{ij})^4\leq |\mS|^{-2} \sum_{j\in\mS^c}
\Big\{ \big(\sum_{i\in\mS}\wt \Psi^2_i\big)\big(\sum_{i\in\mS}\wt\Omega_{ij}^2\big)\Big\}^2 \]
\[\leq  C_{\wt\Psi}\sum_{j\in\mS^c} \Big(\sum_{i\in\mS}\wt\Omega_{ij}^2\Big)^2\leq
C_{\wt \Psi} |\mS|^2 \max_{i\in\mS} \sum_{j\in\mS^c}\wt \Omega_{ij}^4=O(|\mS|^2N^{-\alpha_3})=o(1)\]
for a finite positive constant
$C_{\wt\Psi}$.
Combining the above results, we have proven \eqref{E.1}, which completes the entire proof of the corollary.

\begin{table}[!h]
\renewcommand\arraystretch{1.5}
\bc\emph{}
\caption{\label{tab:tS1} Simulation results with heterogeneous random errors and exogenous covariates.}
\vspace{0.25cm}
{\scriptsize
\begin{tabular}{cccccccccc}
\hline
$N$& TPR & FPR&CFP&Err& &TPR& FPR&CFP&Err  \\
  \hline
 \multicolumn{10}{c}{Example 1: ER model}\\
& \multicolumn{4}{c}{$|\mS_1|=10$} && \multicolumn{4}{c}{$|\mS_1|=15$} \\\cline{2-5}\cline{7-10}
 5000 & 0.969 & 0.000 & 0.540 & 0.192&& 0.955 & 0.000 & 0.550 & 0.156   \\
 7500 & 0.993 & 0.000 & 0.900 & 0.056&& 0.986 & 0.000 & 0.860 & 0.044   \\
 15000& 1.000 & 0.000 & 1.000 & 0.043&& 1.000 & 0.000 & 1.000 & 0.035 \\
 \hline
 \multicolumn{10}{c}{Example 2: stochastic block model}\\
 & \multicolumn{4}{c}{$|\mS_1|=10$} && \multicolumn{4}{c}{$|\mS_1|=15$} \\\cline{2-5}\cline{7-10}
2500 & 0.983 & 0.000 & 0.950 & 0.096 && 0.993 &  0.001 &0.950 & 0.059  \\
5000 & 0.999 & 0.000 & 0.990 & 0.065 && 1.000 &  0.000 &1.000 & 0.044   \\
7500 & 1.000 & 0.000 & 1.000 & 0.049 && 1.000 &  0.000 &1.000 & 0.035 \\
 \hline
 \multicolumn{10}{c}{Example 3: power-law distribution network}\\
 & \multicolumn{4}{c}{$|\mS_1|=10$} && \multicolumn{4}{c}{$|\mS_1|=15$} \\\cline{2-5}\cline{7-10}
5000 &0.968 &0.000 & 0.550 &0.147 && 0.962 & 0.000 & 0.620 & 0.087 \\
10000&0.989 &0.000 & 0.840 &0.071 && 0.985 & 0.000 & 0.850 & 0.050  \\
20000&1.000 &0.000 & 1.000 &0.043 && 1.000 & 0.000 & 1.000 & 0.033 \\
 \hline
\end{tabular}}
\ec
\end{table}
\section{A check on sub-Gaussianity} \label{sec:subgaussion}

 This section contains a check of the Gaussianity. See Fig \ref{fig4} below.
 \begin{figure}
     \centering
     \begin{subfigure}[b]{0.32\textwidth}
         \centering
         \includegraphics[width=\textwidth]{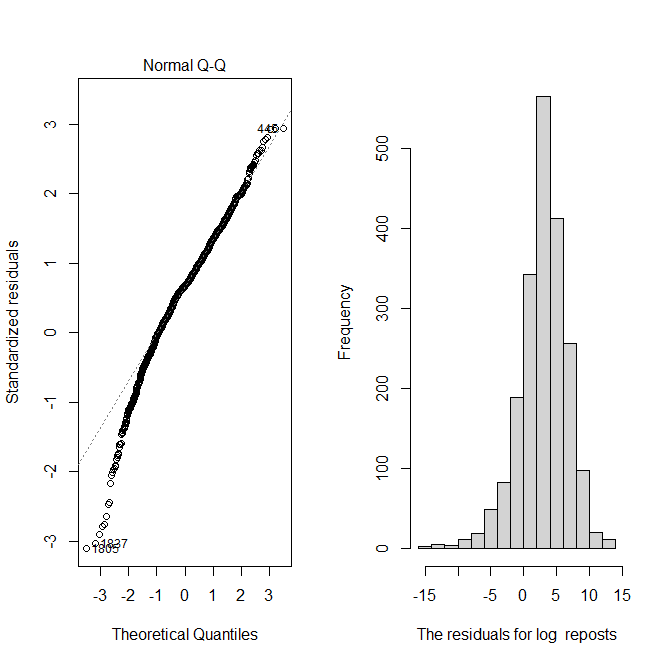}
         \caption{Log Reposts}
     \end{subfigure}
     \hfill
     \begin{subfigure}[b]{0.32\textwidth}
         \centering
         \includegraphics[width=\textwidth]{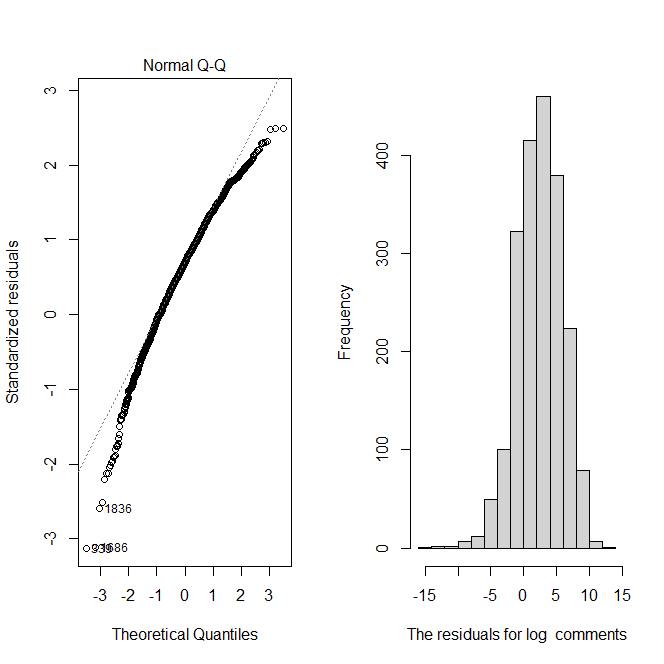}
         \caption{Log Comments}
     \end{subfigure}
     \hfill
     \begin{subfigure}[b]{0.32\textwidth}
         \centering
         \includegraphics[width=\textwidth]{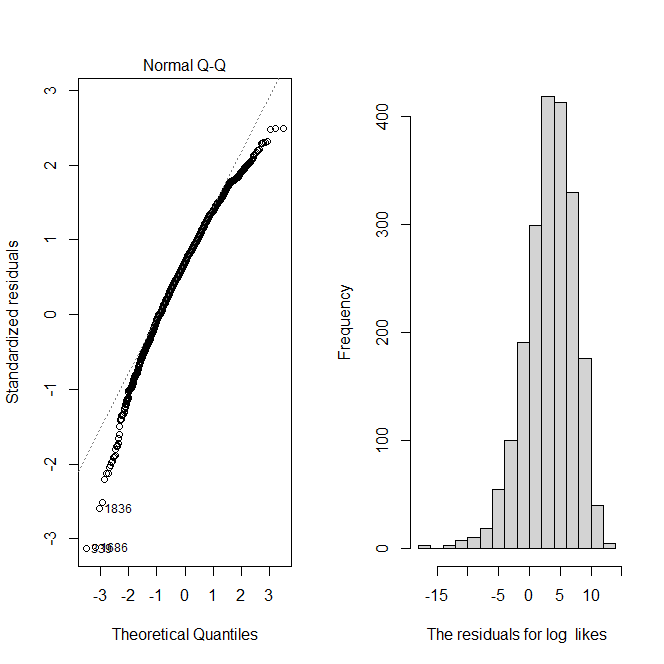}
         \caption{Log Likes}
     \end{subfigure}
        \caption{Empirical distributions of the three responses. }
        \label{fig4}
\end{figure}

\section{Additional simulation results based on data}\label{sec:proportion}
\begin{figure}[htb]
    \centering
    \begin{subfigure}[t]{0.3\linewidth}
        \centering
        \includegraphics[width=\linewidth]{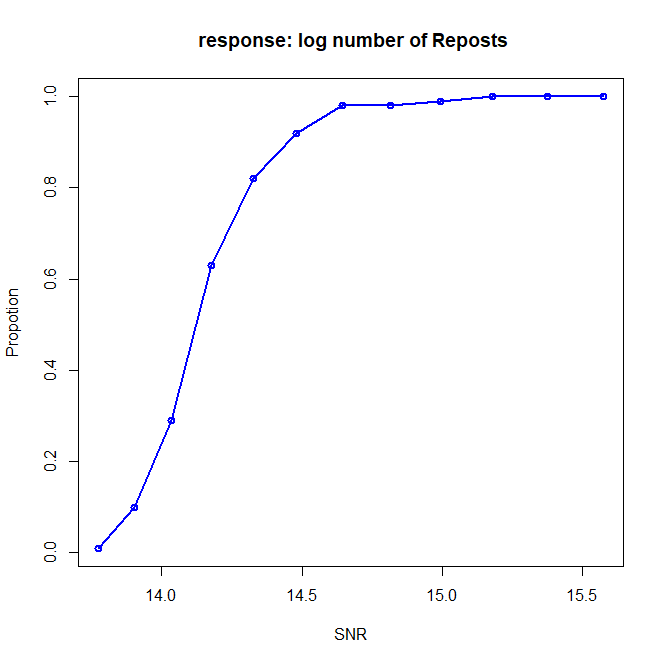}
    \end{subfigure}
    \hfill
    \begin{subfigure}[t]{0.3\linewidth}
        \centering
        \includegraphics[width=\linewidth]{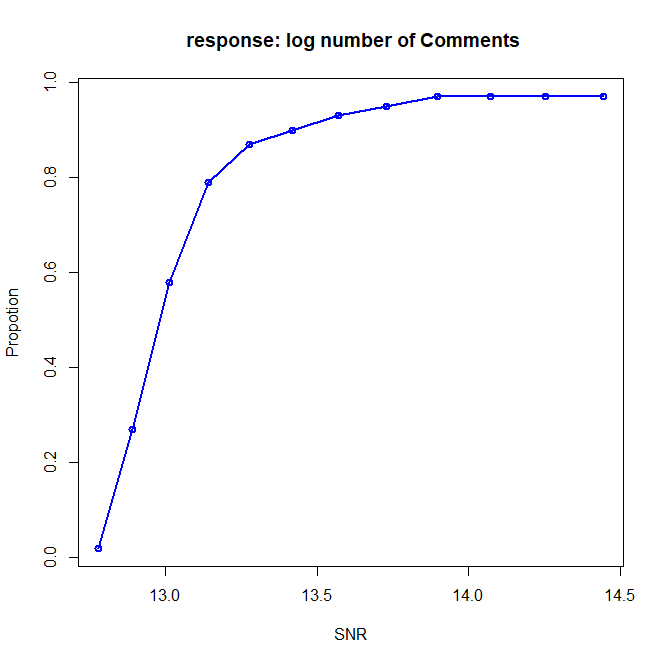}
    \end{subfigure}
    \hfill
    \begin{subfigure}[t]{0.3\linewidth}
        \centering
        \includegraphics[width=\linewidth]{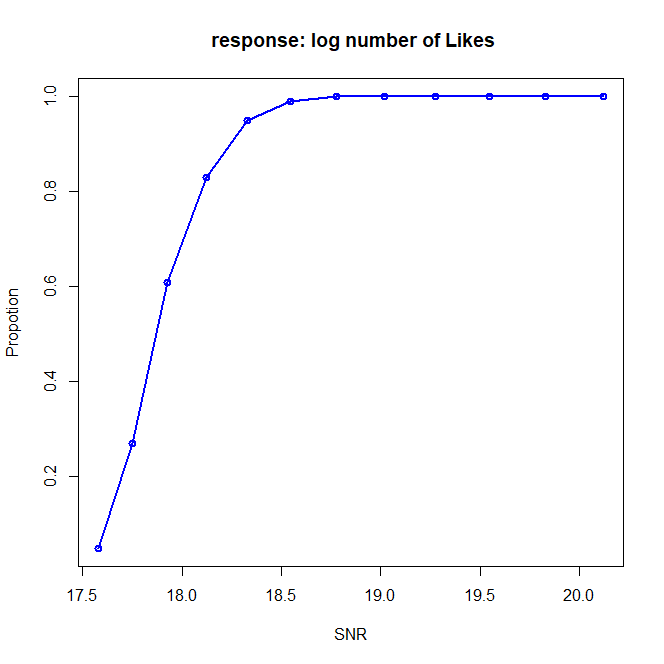}
    \end{subfigure}

    \caption{Proportion of correctly detected influential users as the signal-to-noise ratio (SNR) increases for various response types.}
    \label{fig:proportion_snr}
\end{figure}

This section presents an analysis of the proportion of correctly detected influential users as the signal-to-noise ratio (SNR) increases for different response types. See Fig \ref{fig:proportion_snr} below for the details.

\end{appendix}

\bibliographystyle{agsm}

\bibliography{main}

\end{document}